\documentclass[%
 reprint,
 amsmath,amssymb,
 aps,
]{revtex4-1}

\usepackage{graphicx}
\usepackage{dcolumn}
\usepackage{bm}
\usepackage{color}
\usepackage[T1]{fontenc}
\usepackage[latin1]{inputenc}
\usepackage{amsmath}
\usepackage{amssymb}

\begin{document}

\preprint{APS/123-QED}

\title{Topological and non inertial effects on the interband light absorption}

\author{Moises Rojas}
\email{moises.leyva@dfi.ufla.br}
\author{Cleverson Filgueiras}
 \email{cleverson.filgueiras@dfi.ufla.br}
\affiliation{Departamento de F\'{\i}sica,
Universidade Federal de Lavras, Caixa Postal 3037,
37200-000, Lavras, Minas Gerais, Brazil\\
}%
\author{Julio Brand\~{a}o}
\email{julio.brandao@ifsertao-pe.edu.br}
 \affiliation{Instituto Federal de Educa\c{c}\~ao, Ci\^encia e Tecnologia do Sert\~ao Pernambucano, Campus Salgueiro, 56000-000, Salgueiro, PE, Brazil\\
}%
\author{Fernando Moraes}
\email{fernando.jsmoraes@ufrpe.br}
 \altaffiliation{also at Departamento de F\'{\i}sica, CCEN,  Universidade Federal 
da Para\'{\i}ba, Caixa Postal 5008, 58051-970 , Jo\~ao Pessoa, PB, Brazil }
\affiliation{Departamento de F\'{\i}sica, Universidade Federal Rural de Pernambuco,\\
52171-900, Recife, PE, Brazil\\
 }%

\date{\today}

\begin{abstract}
 In this work, we investigate the combined influence of the nontrivial topology introduced by a disclination and 
non inertial effects due to rotation, in the energy levels and the wave functions of a noninteracting  electron gas confined to a two-dimensional pseudoharmonic quantum dot, under the influence of an external uniform magnetic field. The exact solutions for energy eigenvalues and wave functions are
computed as functions of the applied  magnetic field strength, the disclination topological charge, magnetic quantum number and the rotation speed of the sample. We investigate the modifications on the light interband absorption coefficient and absorption threshold frequency. We observe novel features in the system, including a range of magnetic field without corresponding absorption phenomena, which is due to a tripartite term of the Hamiltonian, involving magnetic field, the topological charge of the defect and the rotation frequency. 
\end{abstract}

\pacs{Valid PACS appear here}
\maketitle


\section{Introduction}

Quantum dots are often referred to as artificial atoms because of their
atom-like electron energy spectrum. They are alluring to a wide range of optoelectronic applications \cite{dotsap}, due to their optical properties, similar to those of atoms \cite{dots1}. They are able to emit light at specific frequencies if either electricity or light is applied to them. The emitted frequency may be precisely tuned by a careful choice of the size of the dot and/or  its shape and composition.  In 
Ref. \cite{PRB.1996.53.6947}, an exactly soluble model to describe quantum
dots, anti-dots, one-dimensional rings and straight two-dimensional wires in
the presence of external fields, was proposed.

An extra ingredient that may influence the tuning of these emission frequencies is  a topological defect \cite{1063-7869-48-7-R02}. Defects, in general, are a nuisance since they impair the electronic properties of the materials. Nevertheless, recent works by some of us \cite{de2012integer,de2013screw,PLA.2012.376.2838,fumeron2017using} have pointed out that topological defects can in fact be used to tailor specific electronic transport properties. Besides the dynamics of carriers, investigations  on how topological defects  affect the electronic bound states, may also be of importance for the improvement of electronic technology. A step forward in this direction was done in Ref. \cite{filgueiras2016landau}.

Rotation, as well, has its effects on quantum systems, like in the celebrated Barnett effect \cite{barnett1915magnetization}.  For instance, in \cite{Lima2014,Lima2015}, the low-energy electronic states of rotating fullerene  were investigated within a continuum model, motivated by the experimental evidence of rapidly rotating $C_{60}$ molecules in fullerite. In Ref. \cite{knutrotation}, it was shown that rotating effects modify the cyclotron frequency and breaks the degeneracy of the analogue  Landau levels for an atom with a magnetic quadrupole moment. The semiclassical kinetic theory of Dirac particles in the presence of external electromagnetic fields and global rotation was established in \cite{PhysRevD.95.085005}. It is clear then, that rotation may be used as an additional tool to manipulate the electronic structure of charge carriers in low dimensional systems as discussed in \cite{EPLrotation,Brandao201555}.
 
As  mentioned above, the interband frequency of absorption/emission by a quantum dot can be tuned with slight changes in parameters like size, shape
and composition, for instance. In this paper, we add two more ingredients: a topological defect called \textit{disclination} and rotation. We investigate how the quantum dots and antidots, with the
pseudoharmonic interaction and under the influence of external magnetic field, are affected by a combination of non inertial and topological influences. Under these circumstances, we obtain exact analytical expressions for the energy spectrum and wavefunctions of a noninteracting two-dimensional electron gas (2DEG) confined in a quantum dot. 
The modifications in the light absorption coefficient is examined and the influences in the threshold frequency value of the absorption coefficient are addressed. Separately, rotation \cite{Brandao201555} and  disclination \cite{Lima2014} couple to the angular momentum, as does the magnetic field. We will see that, when they act together with a magnetic field on a free 2DEG, a new coupling is found involving all of them simultaneously. This makes the energy levels as function of the magnetic field to bend from the usual straight lines and a range of magnetic field without emission/absorption will be observed.

The plan of this work is the following. In Sec. \ref{sec1}, we derive the
Schr\"odinger equation for the 2DEG in a rotating sample, 
with a disclination, in the presence of an external magnetic field and  of a
two-dimensional pseudoharmonic potential. Then, we investigate how such physical conditions affect the electronic energy levels. In reality, we consider the 2DEG confined to a flat interface so that we can discuss our results in the context of a (quasi) two-dimensional electron gas. This permits the inclusion of a disclination, since in three-dimensional systems, the very high elastic energy cost forbids the existence of such a defect. In Sec. \ref{sec2}, we investigate the modifications due to the  topological defect as well as the non inertial effects on
the light interband absorption coefficient and absorption threshold
frequency. The conclusions  are outlined in Section \ref{sec3}.

\section{The Schr\"odinger equation for the system}\label{sec1}
We consider a  2DEG around a disclination, which is a topological defect \cite{1063-7869-48-7-R02} associated to the removal of a wedge of material with the subsequent identification of the loose ends (Volterra process). This introduces an angular deficit, changing the boundary condition on the angular variable from $\phi \rightarrow \phi + 2\pi$ to $\phi \rightarrow \phi + 2\pi\alpha$. Here, $\alpha <1$ expresses the removed wedge angle of $2\pi(1-\alpha)$. Conversely, if a wedge is added, $\alpha >1$. This new boundary condition can effectively be incorporated into the theory if we work out in the background space with  line element
\begin{equation}
ds^2 = dr^2 + \alpha^2 r^2 d\phi^2 +dz^2\;. \label{metric}
\end{equation}
The topological charge of the defect is given by its Frank vector $f$, which  is the curvature flux associated to the defect \cite{1063-7869-48-7-R02}. Since the above line element corresponds to a space with a curvature scalar given by \cite{doi:10.1142/S0217732309029995}
\begin{equation}
R= 2\left( \frac{1-\alpha}{\alpha}\right) \frac{\delta(r)}{r} \;, \label{R}
\end{equation}
its flux is therefore 
\begin{equation}
\oint R r dr d\phi= 4\pi \left( \frac{1-\alpha}{\alpha}\right) = f\;. \label{frank}
\end{equation}
This result still holds for a two-dimensional surface with a disclination, which is the subject of this article.

The Hamiltonian, in cylindrical coordinates, of a charged particle in a disk rotating with angular velocity $\vec{\Omega}=\Omega\hat{z}$, in the presence of a magnetic field $\vec{B}=B \hat{z}$,  can be written as \cite{Brandao201555}
\begin{eqnarray}
H&=&\frac{[\vec{p}-q\vec{A}-m(\vec{\Omega}\times\vec{r})]^2}{2m}-\frac{m(\vec{\Omega}\times\vec{r})^2}{2}+\nonumber\\ &+&qV+V_{ext}\;, \label{hamiltonianageral}
\end{eqnarray}
where $V$ and $\vec{A}$ are the scalar and vector electromagnetic potentials. They are given by
\begin{eqnarray}
V=-\frac{\Omega Br^2}{2},\\
\vec{A}=(0,\frac{Br}{2\alpha},0)\;.
\end{eqnarray}
The electric field associated to the scalar potential appears from the transformation of the applied magnetic field to the rotating frame. The disclination factor, $\alpha$, appearing in the vector potential compensates the change $\phi \rightarrow  \alpha \phi$,  giving the correct value of the magnetic flux through a circle of radius $r$ in the plane. That is, $\oint A_{\phi} ds= \oint \frac{Br}{2\alpha}r \alpha d\phi = \pi r^2 B$,  where $ds = \alpha r d\phi$ comes from the metric (\ref{metric}) with $z=const.$, $r=const.$.
In the same reasoning, we write the angular momentum operator as $\vec{p}_\phi=-\hat{\phi}\frac{i\hbar}{\alpha r}\frac{\partial}{\partial\phi}$.

A scalar pseudoharmonic interaction is incorporated into the system by the potential
\begin{equation}
V_{\mathrm{conf}}=V_{0}\left(\frac{r}{r_{0}}-\frac{r_{0}}{r}%
\right) ^{2},  \label{eq:}
\end{equation}
with $r_{0}=\left( a_{1}/a_{2}\right) ^{1/4}$ and $V_{0}$ being the average radius and the chemical potential \cite{PB.2012.407.4198}, respectively. The choice of the values for the parameters $a_1$ and $a_2$ specifies the particular system under study, as will be seen below. The model given by Eq.  (\ref{eq:}) (see also Ref. \cite{PRB.1996.53.6947,SST.1996.11.1635}) was proposed to describe  quantum dots, anti-dots, one-dimensional rings and
straight two-dimensional wires in the presence of a magnetic field. Some of its properties  are: a) the potential (\ref{eq:}) has a minimum, $V\left( r\right) =0$, at $r=r_{0}$; b) for $r\rightarrow r_{0}$, the potential of the ring has
a parabolic form, $V\left(r\right) \simeq \mu \omega _{0}^{2}\left(
r-r_{0}\right) ^{2}/2$, with $\omega_{0}=\sqrt{8a_{2}/m }$ being the
angular frequency characterizing the strength of the transverse confinement. Another important feature of potential (\ref{eq:}) is that we can control its "shape" such that both the radius and the width of the ring can be adjusted
independently by suitably choosing $a_{1}$ and $a_{2}$. Then, we can study  a $1D$ ring, by choosing $r_{0}=$constant and $\omega _{0}\rightarrow \infty $, a straight 2D wire, by making $\omega _{0}=$constant and $r_{0}\rightarrow\infty $, a quantum dot, by making $a_{1}=0$ and an isolated anti-dot, by making $a_{2}=0$. The particular limits allow us to make a
comparison between the electronic states in different geometries.

Including the above contributions, the Hamiltonian is then written as
\begin{eqnarray}
H=\frac{p^2}{2m}-\mu rp_\phi+\beta r^2 +\frac{V_0r_0^{2}}{r^2}-2V_0\; , \label{H}
\end{eqnarray}
with
\begin{equation}
\mu=\frac{qB}{2m\alpha}+\Omega \label{mu} \\
\end{equation}
and
\begin{eqnarray}
\beta&=&\frac{q^2B^2}{8m\alpha^2}+\frac{qB\Omega}{2}\left(\frac{1-\alpha}{\alpha}\right)+\frac{V_0}{r_0^2}\;.\label{beta}
\end{eqnarray}
The $\mu rp_\phi$ term contains the usual coupling between the magnetic field and the angular momentum. For this Hamiltonian, the Schr\"odinger equation can be written as
\begin{eqnarray}
-\frac{\hbar^2}{2m}\nabla^2\psi&+&i\frac{\mu\hbar}{\alpha}\frac{\partial\psi}{\partial\phi}+\beta r^2\psi+\nonumber\\&+&\left(\frac{V_0r_0^{2}}{r^2}-2V_0\right)\psi=E\psi\;, \label{eqschrodinger}
\end{eqnarray}
where the {\it Laplace-Beltrami} operator is given by $\nabla^2= \frac{\partial^2}{\partial r^2} + \frac{1}{r}\frac{\partial}{\partial r}+ \frac{1}{\alpha^2 r^2}\frac{\partial^2}{\partial \phi^2} $.
With the \textit{ansatz} $\psi=R(r)e^{-il\phi}$ , Eq. (\ref{eqschrodinger}) becomes
\begin{eqnarray}
r^2R''+rR'+\nonumber\\+\left[-\sigma^2r^4+\gamma r^2-\left(\frac{2mV_0r_0^2}{\hbar^2}+\frac{\ell^2}{\alpha^2}\right)\right]R=0, \label{eqemr}
\end{eqnarray}
where $\sigma^2=\frac{q^2B^2}{4\hbar^2\alpha^2}+\frac{mqB\Omega}{\hbar^2}
\left(\frac{1-\alpha}{\alpha}\right)+\frac{2mV_0}{\hbar^2r_0^2}$, 
and $\gamma=\frac{2m}{\hbar}\left(\frac{E+2V_0}{\hbar}-\frac{qB\ell}{2m\alpha^2}-
\frac{\Omega \ell}{\alpha}\right)$. Writing $\sigma r^2=\xi $ and looking at the asymptotic 
limits when $\xi\rightarrow\infty$, the general solution to this equation will be given in terms of  $\rm M(a,b,\xi)$, 
which is the {\it confluent hypergeometric function of the first kind} \cite{Book.1972.Abramowitz},
\begin{eqnarray}
R&=&\nonumber\\
&=&a_{\ell}e^{-\frac{\xi}{2}}\xi^{\frac{|\ell|}{2\alpha}\Lambda}{\rm M}\left(\frac{-\gamma}{4\sigma}+
\frac{|\ell|}{2\alpha}\Lambda+\frac{1}{2} ,1+\frac{|\ell|}{\alpha}
\Lambda,\xi\right)
\nonumber\\
&+&b_{\ell}e^{-\frac{\xi \Lambda}{2\alpha}}\xi^{-\frac{|\ell| \Lambda}{2\alpha}}{\rm M}
\left(\frac{-\gamma}{4\sigma}-\frac{|\ell|}{2\alpha}\Lambda+\frac{1}{2} ,1-\frac{|\ell|}{\alpha}
\Lambda,\xi\right)\;, \nonumber \\
\label{general_sol_2_HO}
\end{eqnarray}
where $\Lambda=\sqrt{1+\frac{2m\alpha^2V_0r_0^2}{\hbar^2|\ell|^2}}$. In Eq. (\ref{general_sol_2_HO}), $a_{\ell}$ and $b_{\ell}$ are, respectively, the coefficients of the {\it regular} and {\it irregular}
solutions. Notice that the term  {\it irregular}  stems from the fact that this solution diverges as $\xi\rightarrow0$. 
We will examine this case since the disclination, described by a cone-like background, introduces a singular
 potential in the problem \cite{Jensen2011448} associated to the curvature given in Eq. (\ref{R}).
In order to have a finite polynomial function 
(the hypergeometric series has to be convergent in order to have a physical solution),
the condition ${\rm a}=-n$, where $n$ is a positive integer number, must be satisfied.
If only the regular solution is to be considered, then $b_{\ell}$ must be zero 
together with the condition $\frac{|\ell|}{\alpha}\sqrt{1+\frac{2m\alpha^2V_0r_0^2}{\hbar^2|\ell|^2}} \geq 1$. Then, the wave function is square integrable.
On the other hand, the constraint $\frac{|\ell|}{\alpha}\sqrt{1+\frac{2m\alpha^2V_0r_0^2}{\hbar^2|\ell|^2}}<1$
must be imposed if the irregular solution is present as a possible wave function. This condition guarantees that it is square
integrable \cite{AoP.2013.339.510,TMP.2009.161.1503,EPJC.2014.74.2708}. 
From these discussions, the discrete possible values for the energy are given by
\begin{eqnarray}
E&=&-2V_0+\frac{\hbar \omega_c \ell}{2 \alpha^2}+
\frac{\hbar\Omega\ell}{\alpha}+\nonumber\\&+&\hbar\sqrt{ \frac{\omega_{c}^2}{\alpha^2}+
4\omega_c\Omega\frac{\left(1-\alpha\right)}{\alpha}+
\frac{8V_0}{mr_0^2}}\times\nonumber \\ 
& &\left[n\pm\frac{1}{2}\sqrt{\frac{\ell^2}{\alpha^2}+
\frac{2mV_0r_0^2}{\hbar^2}}+\frac{1}{2}\right], \label{Energyspectrum}
\end{eqnarray}
where $\omega_c=qB/m$ is the cyclotron frequency. The ``$+$'' sign stands for the regular 
wave solution while the ``$-$'' sign stands for the irregular one. 
$\vec{B}$ and $\vec{\Omega}$ are tunable external parameters. 
Considering $\alpha=1$ (flat), $V_0=0$ and $\Omega\neq 0$, we recover the results found in Ref. \cite{Brandao201555}. By considering 
$\alpha\neq 1$, $V_0=0$ and $\Omega=0$, we recover the results found in Ref. \cite{deLima2012}. 
\section{Interband light absorption}\label{sec2}
In this section, we calculate the direct interband light absorption coefficient $K(\omega )$ and the absorption threshold frequency  in a
quantum pseudodot under the simultaneous influence of an external magnetic field, rotation and disclination.
The light absorption coefficient can be expressed as \cite{PE.2004.22.860,PE.2006.31.83,PB.2005.363.262,SSS.2010.12.1253}
\begin{align}
K(\omega )& =N  \notag \\
& \times \sum_{n,l,\mu }\sum_{n^{\prime },l^{\prime },\mu ^{\prime
}}\left\vert \int \psi _{n,l,\mu }^{e}\left( r ,\phi \right) \psi
_{n^{\prime },l^{\prime },\mu ^{\prime }}^{h}\left( r ,\phi \right) r
dr d\phi \right\vert ^{2}  \notag \\
& \times \delta \left( \Delta -E_{n,l,\mu }^{e}-E_{n^{\prime },l^{\prime
},\mu ^{\prime }}^{h}\right)\;,  \label{absor}
\end{align}%
where
\begin{eqnarray}
\mu\equiv\pm\sqrt{\frac{|\ell|^2}{\alpha^2}+\frac{2m_{e}V_0r_0^2}{\hbar^2}}\;,
\end{eqnarray}

\begin{eqnarray}
\mu^{\prime}\equiv\pm\sqrt{\frac{|\ell|^2}{\alpha^2}+\frac{2m_{h}V_0r_0^2}{\hbar^2}}\;,
\end{eqnarray}
$\Delta \equiv \hbar \varpi-\varepsilon _{g}$, $\varepsilon _{g}$ is the width of the forbidden energy gap,
$\varpi$ is the frequency of incident light, $N$ is a quantity proportional to the square of dipole
moment matrix element modulus, $\psi ^{e\left( h\right) }$ is the wave
function of the electron (hole) and $E^{e(h)}$ is the corresponding energy of
the electron (hole). Considering the solution (\ref{general_sol_2_HO}),  Eq.(\ref
{absor}) becomes \cite{PB.2012.407.4198}
\begin{align}
K(\omega )& =N\sum_{n,l,\mu }\sum_{n^{\prime },l^{\prime },\mu ^{\prime }}
\frac{\sigma ^{ \mu  + \mu ^{\prime
} +2}(n+\mu  )!(n^{\prime }+ \mu
^{\prime } )!}{\pi ^{2}n!n^{\prime }!\left( \left\vert \mu
\right\vert !\right) ^{2}(\left\vert \mu ^{\prime }\right\vert !)^{2}}
\notag \\
& \times \Bigg\vert\int_{0}^{2}\pi e^{i(l+l^{\prime })}\int_{0}^{\infty
}r dr e^{-\frac{1}{2}(\sigma +\sigma ^{^{\prime }})r ^{2}}r
^{ \mu  +\mu ^{\prime } }  \notag
\\
& \times \,\mathrm{M}\left( -n,1+ \mu  ,\sigma r
^{2}\right) \mathrm{M}\left( -n^{\prime },1+ \mu ^{\prime
},\sigma ^{\prime }r^{2}\right) \Bigg\vert^{2}  \notag \\
& \times \,\delta \left( \Delta -E_{n,l,\mu }^{e}-E_{n^{\prime },l^{\prime
},\mu ^{\prime }}^{h}\right)\;.  \label{absor2}
\end{align}%

Following Ref. \cite{PB.2012.407.4198}, the light absorption coefficient is given by
\begin{equation}
K\left( \omega \right) =N\sum_{n,l,\mu}\sum_{n^{\prime},l^{\prime},\mu
^{\prime }}P_{nn^{\prime}}^{\mu}Q_{nn^{\prime }}^{\mu}\delta \left(
\Delta -E_{n,l,\mu}^{e}-E_{n^{\prime},l^{\prime},\mu^{\prime
}}^{h}\right)\;,  \label{absor3}
\end{equation}%
where
\begin{align*}
P_{nn^{^{\prime}}}^{\mu}& =\frac{1}{\left(\mu 
!\right)^{4}}(\sigma \sigma^{\prime})^{ \mu 
+1}\left(\frac{\sigma+\sigma^{\prime}}{\sigma-\sigma^{\prime}}\right)
^{2\left(n+n^{\prime}\right)} \\
& \times \frac{\left( n+\mu  \right) !\left(n^{\prime
}+ \mu  \right) !}{n!n^{\prime}!}\;,
\end{align*}
and
\begin{equation*}
Q_{nn^{\prime }}^{\mu }=\left[ A_{ \mu  ,\sigma }
\mathrm{M}\left( n,n^{\prime }, \mu  +1;-\frac{4\sigma
\sigma ^{\prime }}{\left( \sigma -\sigma ^{\prime }\right) ^{2}}\right)
\right] ^{2},
\end{equation*}
where
\begin{equation*}
A_{ \mu  ,\sigma}=\left(\lvert \mu \rvert
\right) !\left( \frac{2}{\sigma +\sigma ^{\prime }}\right) ^{2 \mu
 +1}\;.
\end{equation*}

The argument of the Dirac delta function in Eq. (\ref{absor3}) together with Eq. (\ref{Energyspectrum}), allows us to define the absorption  threshold frequency  as
\begin{eqnarray}
\hbar\varpi&=&\varepsilon_g+\left[\hbar\left(n+\frac{\mu}{2}+\frac{1}{2}\right)\Xi_{e}+
\frac{\hbar\ell}{2\alpha}\left(-\frac{\omega_{ce}}{\alpha}+2\Omega\right)\right]+\nonumber\\&+&
\left[\hbar\left(n^{\prime}+\frac{\mu^{\prime}}{2}+\frac{1}{2}\right)\Xi_{h}+
\frac{\hbar\ell^{\prime}}{2\alpha}\left(\frac{\omega_{ch}}{\alpha}+2\Omega\right)\right]-4V_0, \nonumber \\
\end{eqnarray}
where
\begin{equation}
\Xi_{e}=\sqrt{\frac{\omega_{ce}^{2}}{\alpha^{2}}-
4\omega_{ce}\Omega\frac{(1-\alpha)}{\alpha}+\frac{8V_{0}}{m_{e}r_{0}^{2}}}\;, \label{22}
\end{equation}
for electrons, and 
\begin{equation}
\Xi_{h}=\sqrt{\frac{\omega_{ch}^{2}}{\alpha^{2}}+
4\omega_{ch}\Omega\frac{(1-\alpha)}{\alpha}+\frac{8V_{0}}{m_{h}r_{0}^{2}}}\;, \label{23}
\end{equation}
for holes. In these equations, $\omega_{ce}=eB/m_e$ and $\omega_{ch}=eB/m_h$, where $e$ is the elementary charge and $m_e$ and $m_h$ are, respectively, the effective masses of the electron and the hole.

Let us now investigate the influence a disclination ($\alpha$) as well as  rotation ($\Omega$) on the
threshold value of the absorption for the transition $00\rightarrow 00$, comparing
with the case where these ingredients are absent. The threshold value of absorption is then given by
\begin{eqnarray}
\hbar\varpi_{00}&=&\varepsilon_g+\left[\hbar\left(\pm\sqrt{\frac{m_{e}V_0r_0^2}{2\hbar^2}}+\frac{1}{2}\right)\Xi_{e}\right]\nonumber\\&+&\left[\hbar\left(\pm\sqrt{\frac{m_{h}V_0r_0^2}{2\hbar^2}}+\frac{1}{2}\right)\Xi_{h}\right]-4V_0.\label{zero}
\end{eqnarray}

From Eq. (\ref{zero}), we can define the dimensionless absorption threshold  as 

\begin{equation}
W=\frac{\hbar\varpi_{00}}{\varepsilon_{g}}. \label{W}
\end{equation}
For a quantum dot $(V_{0}r_{0}^{2}\rightarrow0)$, we obtain the absorption threshold
frequency  as

\begin{equation}
W=1+\frac{\hbar}{2\varepsilon_{g}}(\Xi_{e}+\Xi_{h})-\frac{4V_{0}}{\varepsilon_{g}}. \label{26}
\end{equation}

Before analyzing the consequences of Eq. (\ref{26}) we note that the second term in the square root in Eqs. (\ref{22}) and (\ref{23}) is proportional to the product $eB\Omega f$, where the topological charge $f$ is given by Eq. (\ref{frank}). This remarkable term shows how the electric charge of the carrier, the magnetic field strength, the rotation speed and the topological charge of the disclination couple as a simple product. This has profound implications in the absorption threshold since it  is sensitive to the signs of each one of them. So, whether the carrier is an electron or a hole, or whether $\vec{B}$ and $\vec{\Omega}$ are parallel or anti-parallel,   or whether the disclination has negative or positive curvature, it all matters. Whatever the combination of signs entering the product, the sign in front of the corresponding terms in Eqs. (\ref{22}) and (\ref{23}) will always be opposite. This implies that, in some cases, either equation might yield  imaginary results which indicate that there are no bound states in these conditions (see Eq. (\ref{Energyspectrum})). This explains the sudden stop of two of the curves in Fig. \ref{fig:fvsr0} below.

In Fig. \ref{fig:fvsB1}, we show the absorption threshold  frequency 
$W$ versus magnetic field $B$, for different values of $\alpha$
and of $\Omega$. In Fig. \ref{fig:fvsB1}(a) it can
be observed that for any disclination $\alpha$ and angular frequency
$\Omega$, the behavior is linear for strong applied magnetic fields.
One  can also see that, for a positive curvature
disclination ($\alpha<1$), the frequency increases faster than in the negative curvature disclination ($\alpha>1$) case. On the other hand, in
Fig. \ref{fig:fvsB1}(b) one can see that, for weak magnetic fields,
the absorption threshold  frequency  is nonlinear. Furthermore, it is seen  that an increase of the angular frequency from $\Omega=0$ (red
dashed line) to $\Omega=500$ Ghz (red solid line) rises the absorption
threshold  frequency  $W$ for $\alpha=\frac{5}{6}$ and
$B>0$, while for $B<0$, the effect is opposite. Also, we can see
that, increasing the angular frequency $\Omega=0$ (blue dashed line) to
$\Omega=500$ Ghz (blue solid line), lowers the absorption threshold  frequency
 for $\alpha=\frac{7}{6}$ and $B>0$. In the
region $B<0$ the effect is opposite. Finally, when the system
has no disclination ($\alpha=1$), the threshold frequency is independent
of $\Omega$ (the solid and the dashed line coincide).

\begin{figure}
\includegraphics[scale=0.35]{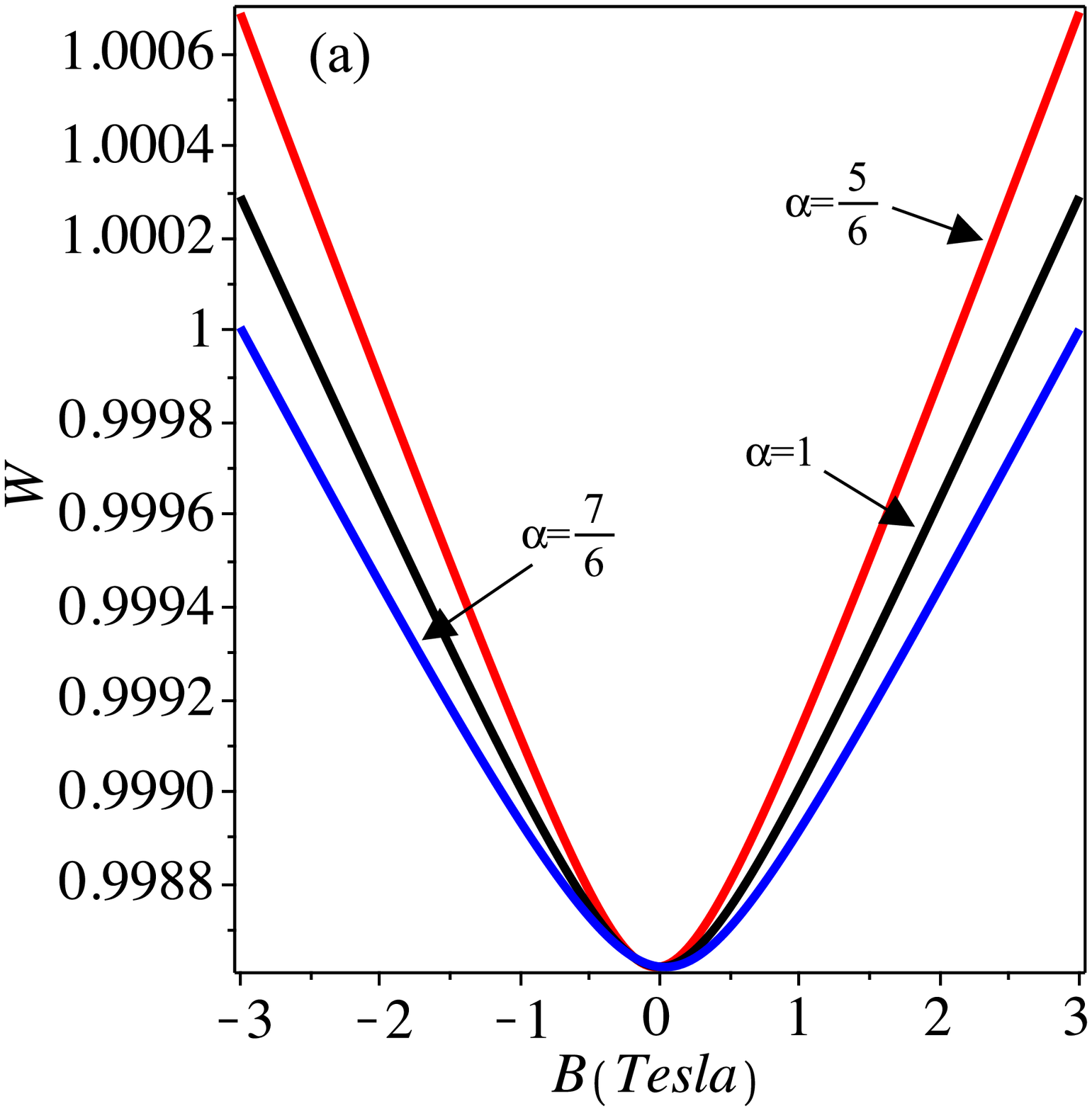}\\ \includegraphics[scale=0.37]{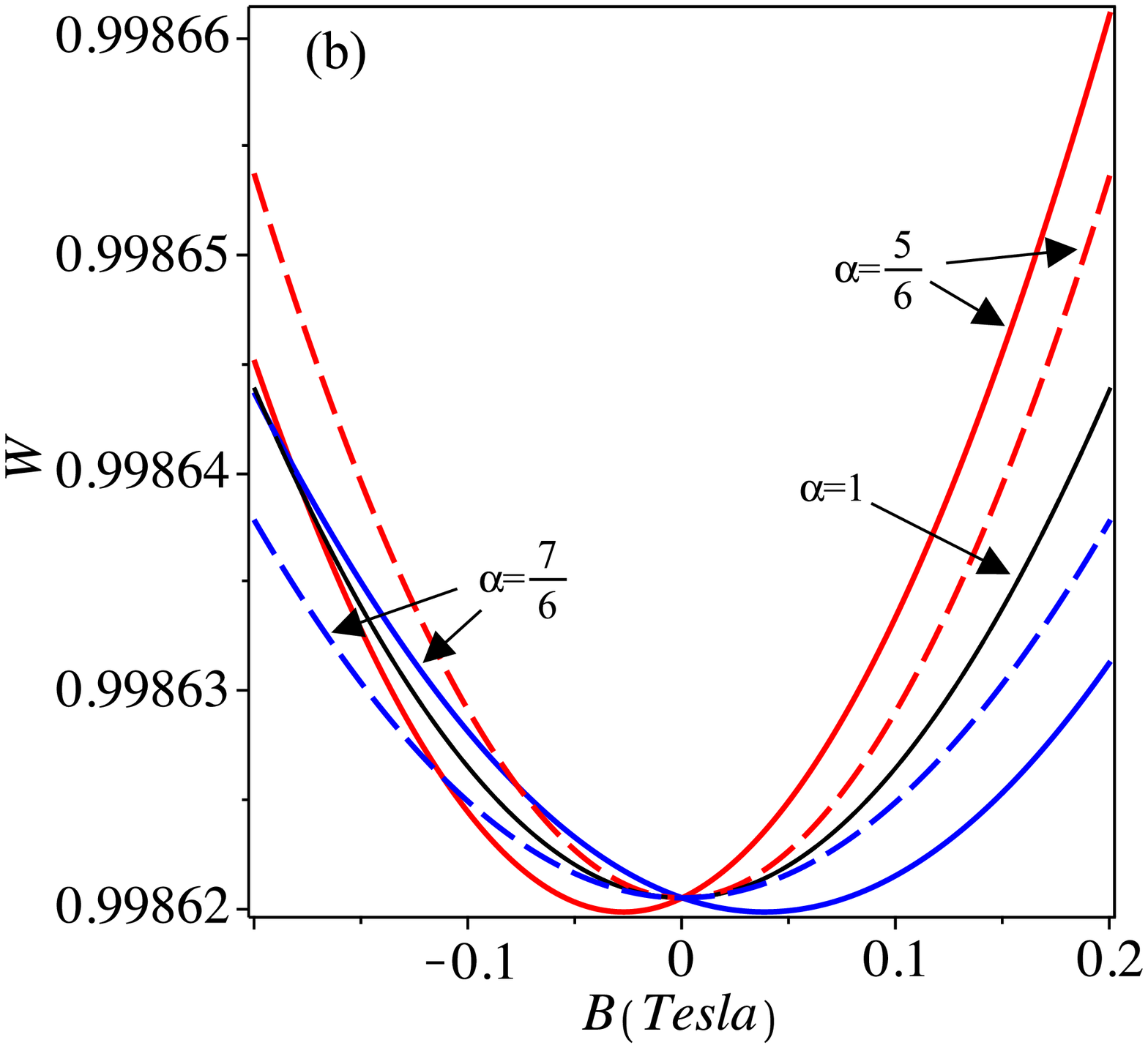}\caption{\label{fig:fvsB1}(Color online) The absorption threshold  frequency 
$W$ as a function of magnetic field $B$ for $\Omega=500$ Ghz (solid lines),
$\Omega=0$ Ghz (dashed lines) and three values of the disclination parameter $\alpha$: (a)
strong magnetic field  and (b) weak magnetic field. The quantum dot  size $r_0 = 895.8$ $\mathring{A}$ was assumed for all plots.}

\end{figure}
\begin{figure}

\includegraphics[scale=0.34]{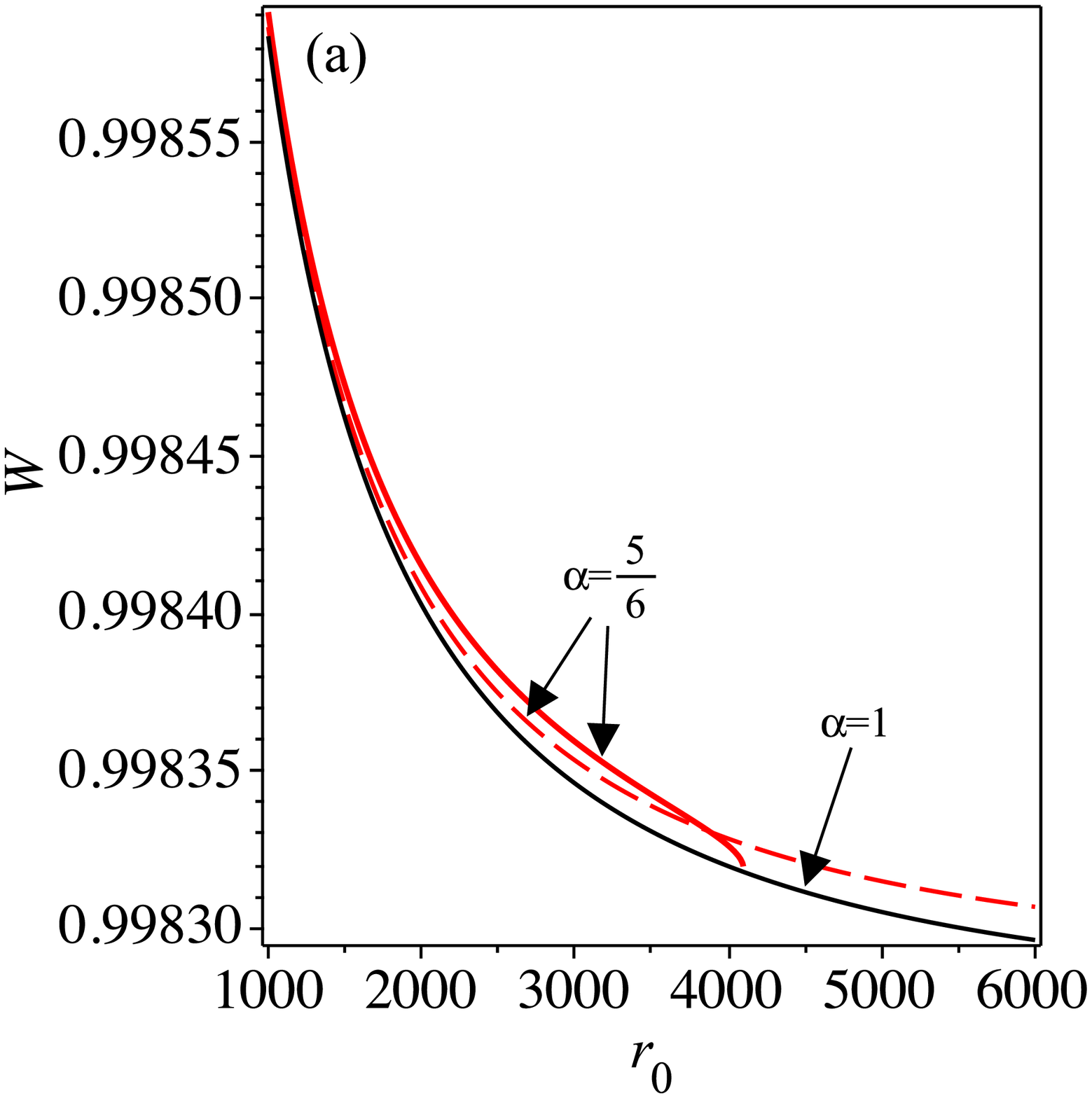}\\
\includegraphics[scale=0.34]{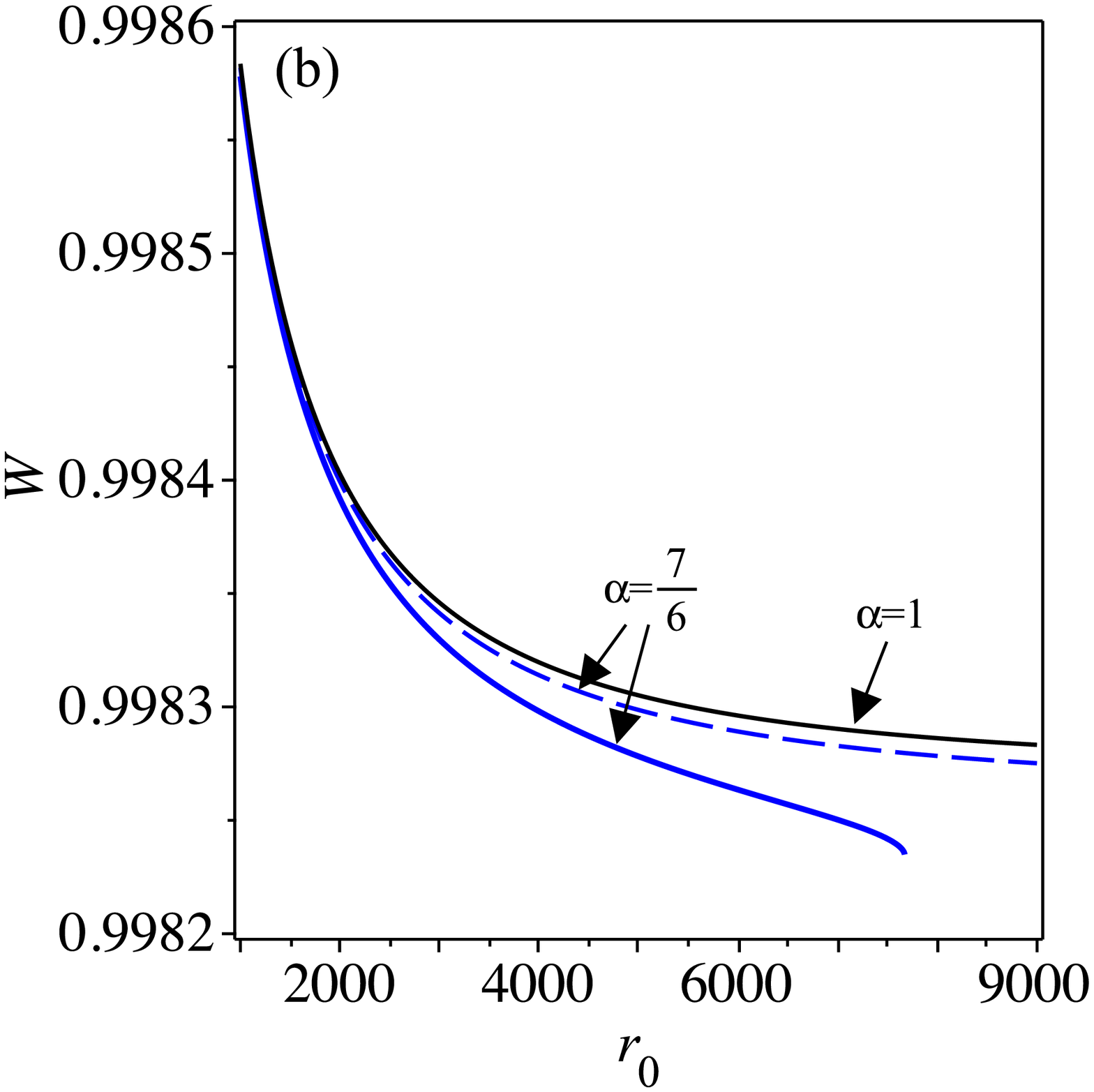}\caption{\label{fig:fvsr0}(Color online) The absorption threshold  frequency $W$ versus quantum dot size $r_0$ (in  $\mathring{A}$)  for $B=0.1T$ (all lines) and   $\Omega=0$ (dashed lines), $\Omega=500$ Ghz
(solid lines): (a)  $\alpha = 5/6$ and  (b) $\alpha = 7/6$.  The $\alpha = 1$, or defectless case (solid black line), which does not depend on $\Omega$, is also plotted  for comparison.  Notice the abrupt ending of the colored solid lines in both cases.}

\end{figure}

In Fig. \ref{fig:fvsr0}, we present  plots of the absorption threshold  frequency
 as a function of the quantum dot size $r_{0}$ (in  $\mathring{A}$) for  $\Omega=0$, $\Omega=500$ Ghz, $B=0.1T$ and a few values of the disclination parameter $\alpha$. As  seen
from the figures, the absorption threshold  frequency  decreases with the
increasing of the quantum dot radius in all cases. In Fig. \ref{fig:fvsr0}(a), we note that the threshold frequency
decreases monotonically for $\alpha=1$ (black solid line), $\alpha=\frac{5}{6}$
(red solid line), both of these at $\Omega=500$ GHz. Same behavior for the  $\Omega=0$ case. Moreover, we find that the threshold frequency
decreases monotonically but suddenly disappears in the case of the
disclination assuming the value $\alpha=\frac{5}{6}$ (red solid line)
and $\Omega=500$ GHz. Fig. \ref{fig:fvsr0}(b) shows that the curve
of the absorption threshold  frequency shifts down when the
disclination parameter changes from $\alpha=1$ (black solid line) to $\alpha=\frac{7}{6}$
(blue dashed line). One interesting observation is that the absorption threshold  frequency
 suddenly disappears for disclination parameter $\alpha=\frac{7}{6}$
(blue solid line) and $\Omega=500$ GHz. As mentioned above, the abrupt ending of the curves in Fig. \ref{fig:fvsr0} is due to the absence of bound states in the region beyond. That is, for a given disclination, for fixed values of $B$ and $\Omega$, there is an upper limit for the radius of the quantum dot. This is clear if one takes a look at the Hamiltonian  (\ref{H}). If $\beta$, as given by Eq. (\ref{beta}), becomes negative, clearly, the effective potential in  (\ref{H}) does not yield bound states.

\begin{figure}
\includegraphics[scale=0.42]{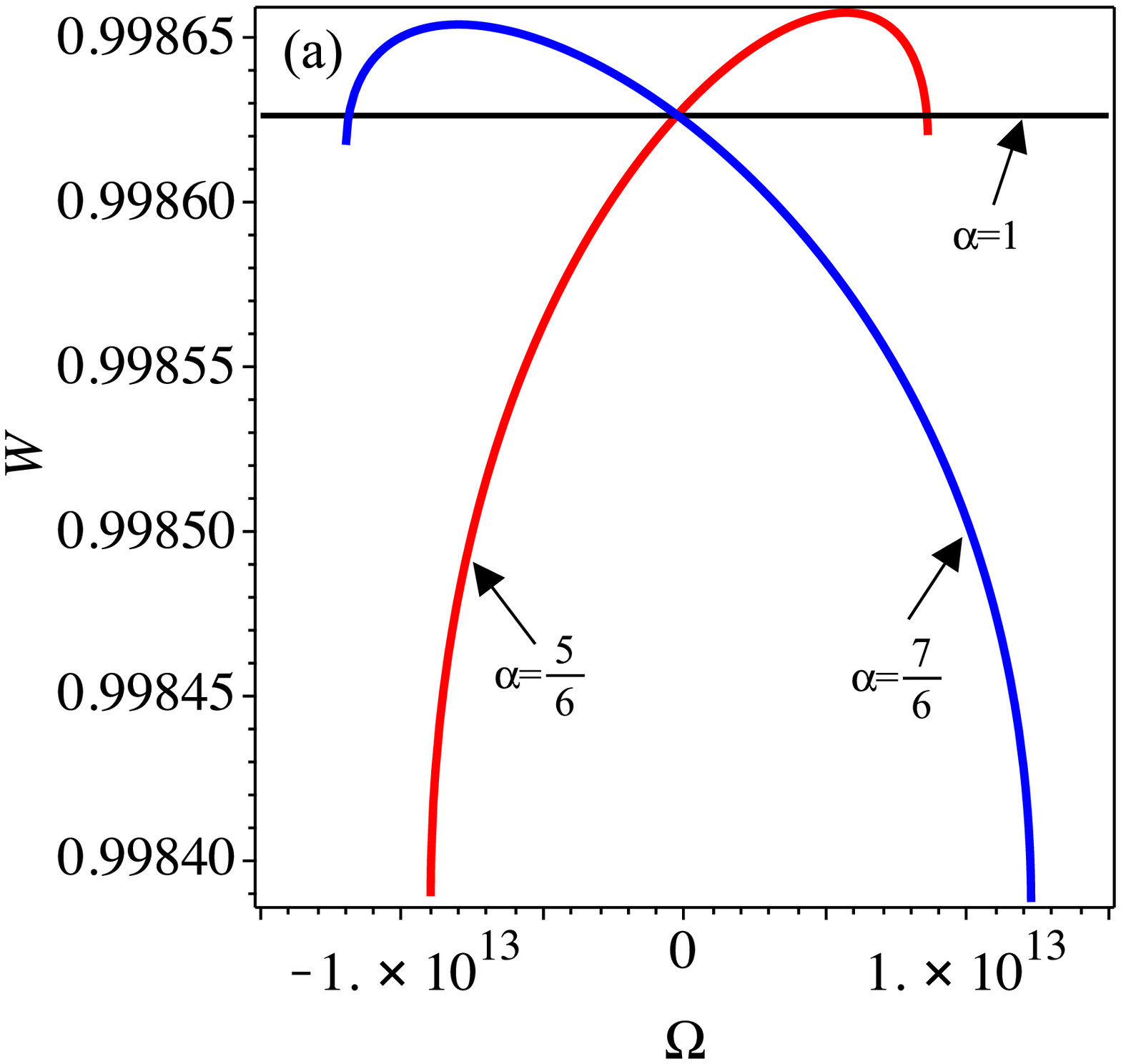}
\\
\includegraphics[scale=0.36]{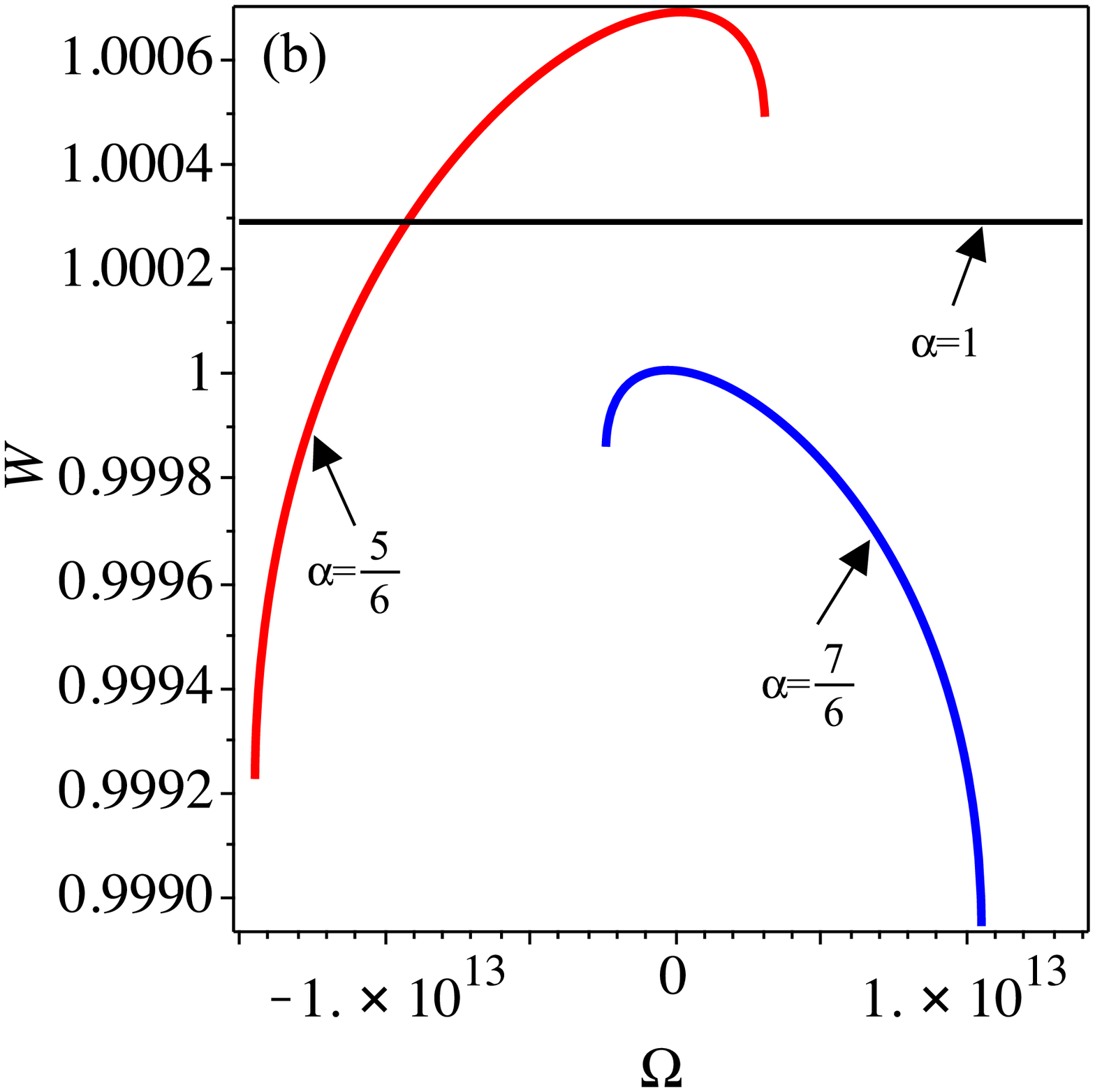}\caption{\label{fig:fvsOmega}(Color online) The absorption threshold   frequency $W$ as a function of angular frequency $\Omega$ for $\alpha=\frac{5}{6}$,
$\alpha=1$, $\alpha=\frac{7}{6}$. (a) $B=0.1 T$. (b) $B=3 T$. The quantum dot  size $r_0 = 895.8$ $\mathring{A}$ was assumed for all plots.}

\end{figure}

In Fig. \ref{fig:fvsOmega}, we investigate the effect of the angular
frequency $\Omega$ on the absorption threshold  frequency $W$
for magnetic field $B=0.1T$ and $B=3T $, and also for three different values of the
disclination parameter $\alpha$. First, it can be seen from Fig.
\ref{fig:fvsOmega}(a), where $B=0.1T$, that when $\alpha=\frac{5}{6}$ (a wedge is
removed), the maximum of the absorption frequency  corresponds to an anticlockwise (positive)
angular velocity $\Omega$ and it decreases with increasing $\Omega$ until it suddenly disappears. Furthermore, 
for clockwise (negative) angular velocity, the absorption frequency  continues
to decrease as $|\Omega|$ increases. In the case where $\alpha=\frac{7}{6}$ (a wedge is added),
the figure shows a nearly symmetric behavior with the threshold frequency increasing when anticlockwise angular
velocity  decreases. On the other hand, in the region of clockwise angular
velocity  the threshold frequency increases with increasing $|\Omega|$
until it reaches the maximum peak. Soon after this, it suddenly disappears.
Finally, when the system has no disclination, $\alpha=1$, the threshold
frequency absorption is independent of the angular frequency. 

As shown in Fig. \ref{fig:fvsOmega}(b) for $B=3T$, the gap between the threshold frequency maximum  for $\alpha=\frac{5}{6}$ and $\alpha=\frac{7}{6}$
is greater than in case $B=0.1T$. The stronger magnetic field has moved the $W(\Omega)$ curves with respect to each other as compared to  Fig. \ref{fig:fvsOmega}. Now the $\alpha=\frac{5}{6}$ and $\alpha=\frac{7}{6}$ no longer cross each other. 

When we consider the quantum anti-dot $\left(\frac{V_{0}}{r_{0}^{2}}\rightarrow0\right)$,
the absorption threshold  frequency  is given by the expression

\begin{eqnarray}
W=1-\frac{4V_{0}}{\varepsilon_{g}}+\frac{\hbar}{2\varepsilon_{g}}\left[\Phi_{e}\left(1+\sqrt{\frac{2m_{e}V_{0}r_{0}^{2}}{\hbar^{2}}}\right)+\right.\nonumber\\ \left.+\Phi_{h}\left(1+\sqrt{\frac{2m_{h}V_{0}r_{0}^{2}}{\hbar^{2}}}\right)\right]\;,
\end{eqnarray}
where $\Phi_{e(h)}=\sqrt{\frac{\omega_{ce(h)}^{2}}{\alpha^{2}}\pm4\omega_{ce(h)}\Omega\frac{(1-\alpha)}{\alpha}}$.

\begin{figure}
\includegraphics[scale=0.33]{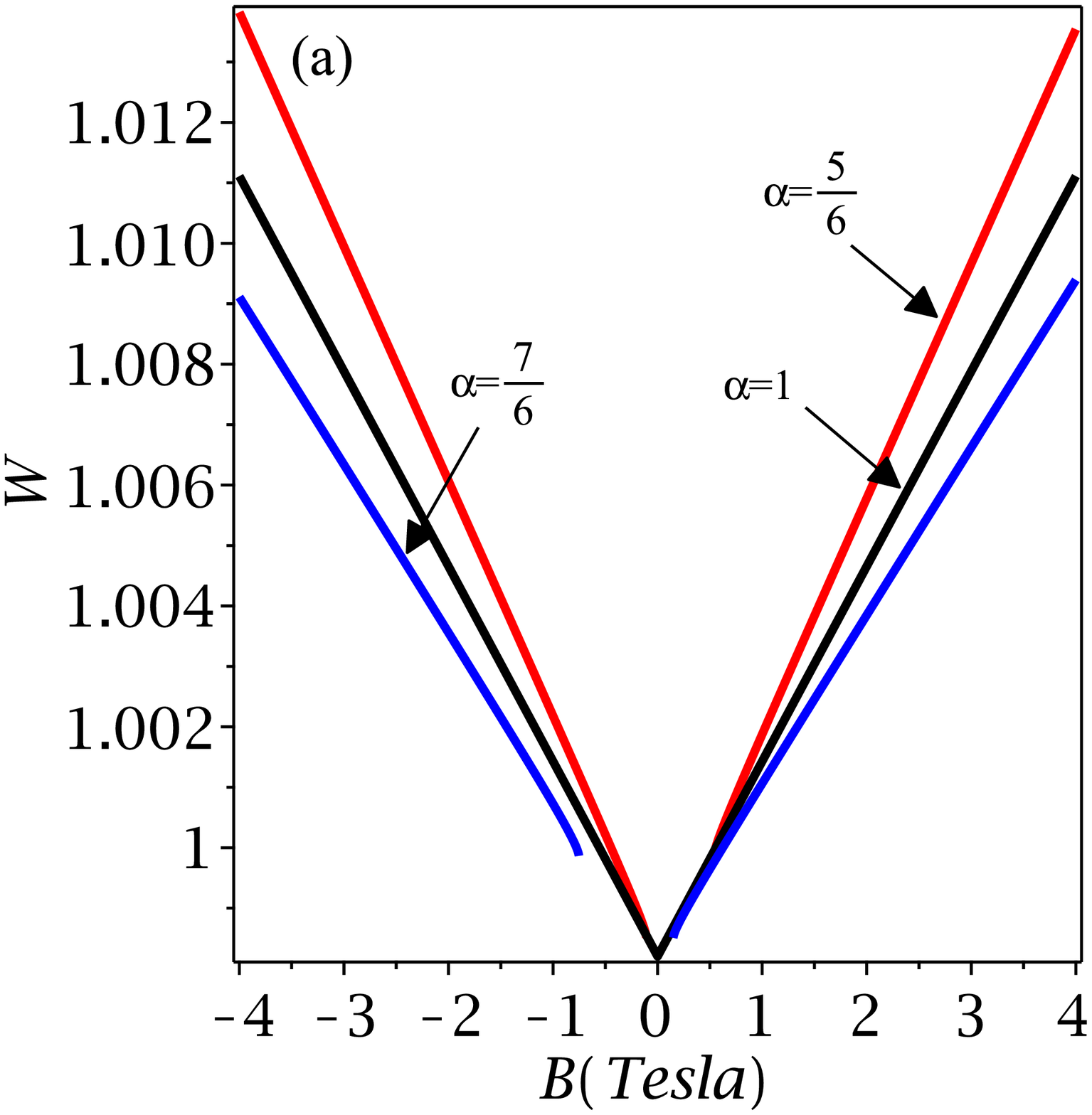}\\ \includegraphics[scale=0.33]{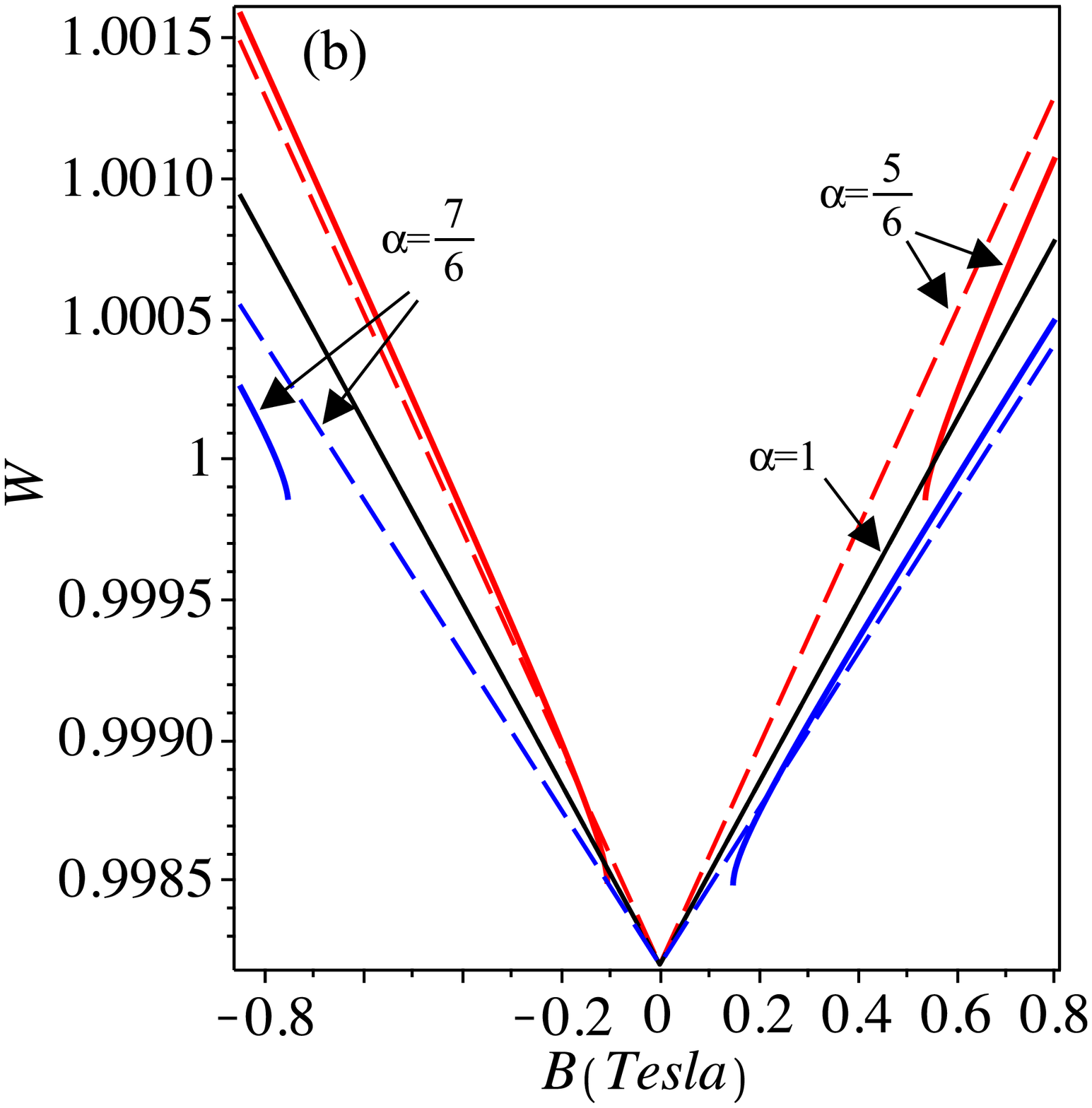}\caption{\label{fig:fvsB-anti}(Color online) The absorption threshold  frequency   $W$ for a quantum anti-dot as a function of magnetic field $B$, for several
values of the disclination parameter $\alpha$ and for $\Omega=500$ Ghz. (a) strong
magnetic field. (b) weak magnetic field. The quantum anti-dot  size $r_0 = 895.8$ $\mathring{A}$ was assumed for all plots.}

\end{figure}

Fig. \ref{fig:fvsB-anti} shows the variation of the absorption threshold 
frequency  $W$ for a quantum anti-dot as a function of the
magnetic field $B$, for fixed $\Omega=0$, $\Omega=500$ Ghz and
different values of the disclination parameter $\alpha$. As illustrated
in Fig. \ref{fig:fvsB-anti}(a) the behavior of the threshold frequency
is linear for strong magnetic fields for $\alpha=\frac{5}{6}$ (red
solid line), $\alpha=1$ (black solid line), $\alpha=\frac{7}{6}$
(blue solid line) and independent of the values of $\Omega$. On the
other hand, in Fig. \ref{fig:fvsB-anti}(b), we show the magnetic
field dependence of the threshold frequency for small magnetic fields.
The results show clearly that for $\Omega=0$, the behavior of the
threshold frequency, for several values of parameter $\alpha$, is linear.
However, the most striking finding happens when angular velocity is
not null (here, we consider $\Omega=500$ Ghz) for $\alpha=\frac{5}{6}$
(red solid lines) and $\alpha=\frac{7}{6}$ (blue solid lines) it
is found that, in the weak magnetic field region 
the threshold frequency suddenly disappears. Again, the reason is the absence of bound states due an imaginary value of $\beta$.

\begin{figure}
\includegraphics[scale=0.32]{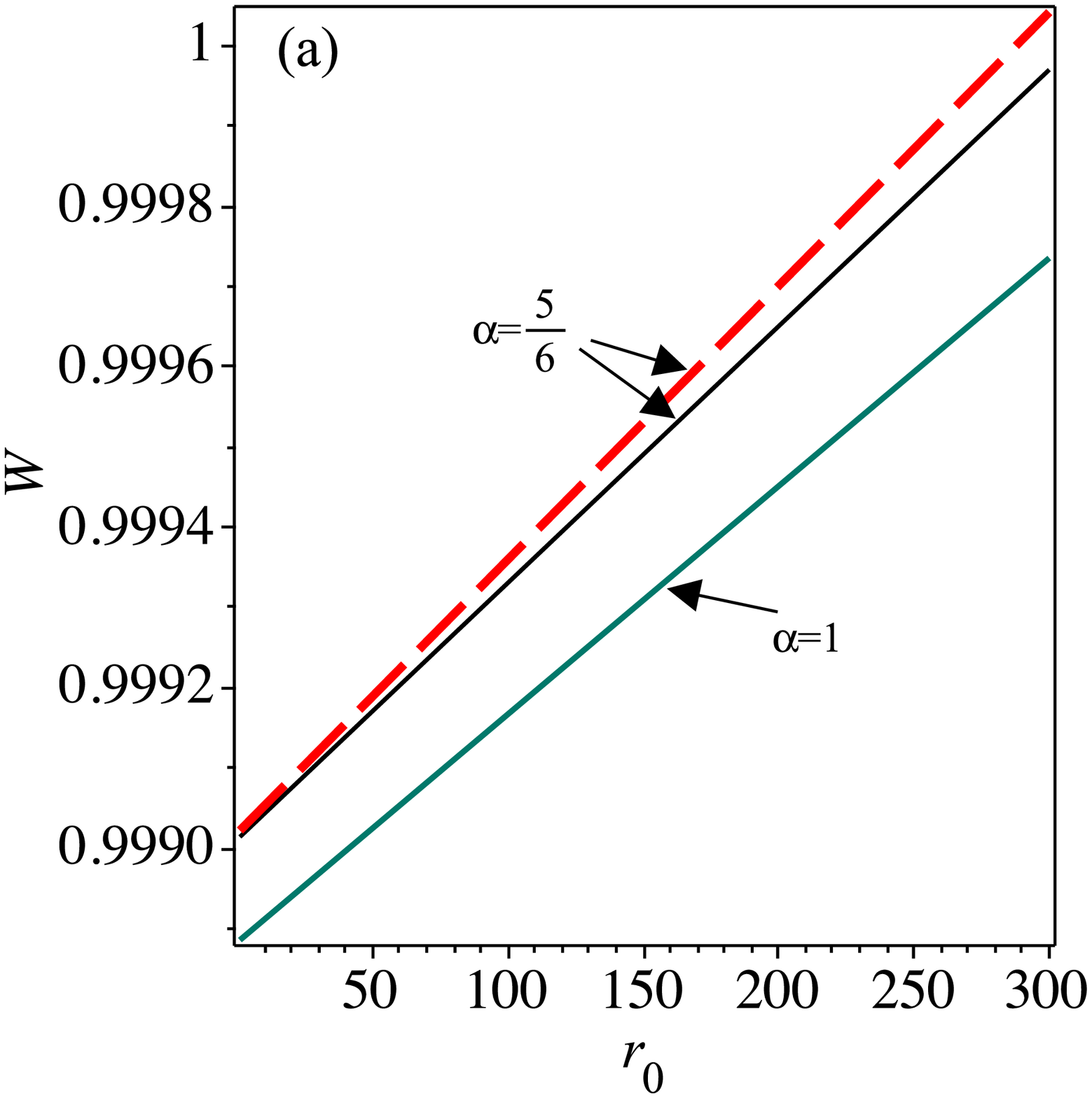}\\ \includegraphics[scale=0.32]{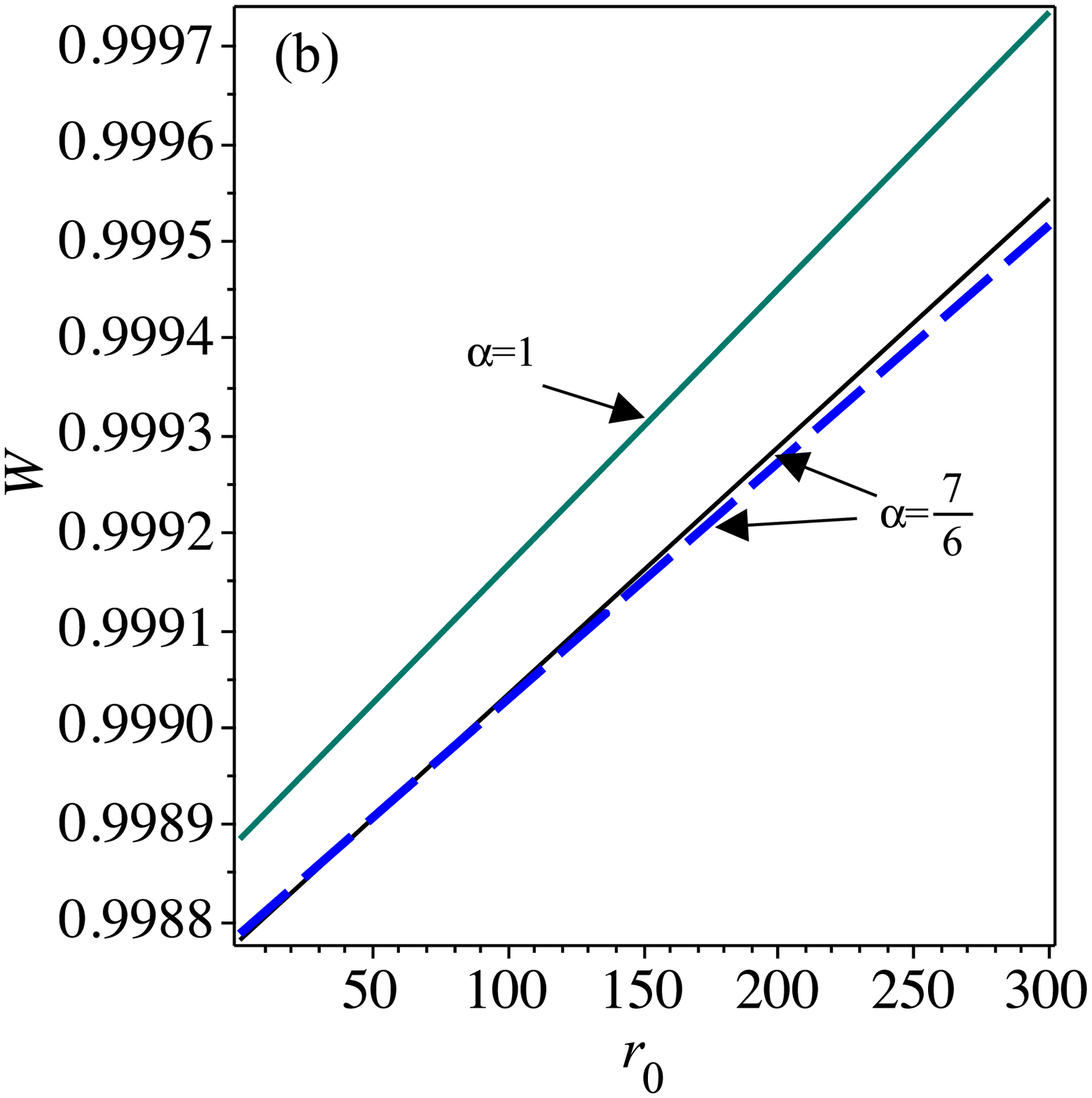}\caption{\label{fig:fvsr}(Color online) The absorption threshold  frequency 
$W$ versus quantum anti-dot size $r$ for the fixed value $B=1 T$.
(a) For $\alpha=\frac{5}{6}$ with $\Omega=500$ GHz (black line),
$\Omega=0$ (red dashed line) and $\alpha=1$ with $\Omega=500$ GHz
(green line). (b) For $\alpha=\frac{7}{6}$ with $\Omega=500$ GHz
(black line), $\Omega=0$ (blue dashed line). The green line corresponds
to $\alpha=1$ with $\Omega=500$ GHz. }
\end{figure}

\vspace{5mm}

\begin{figure}
\includegraphics[scale=0.33]{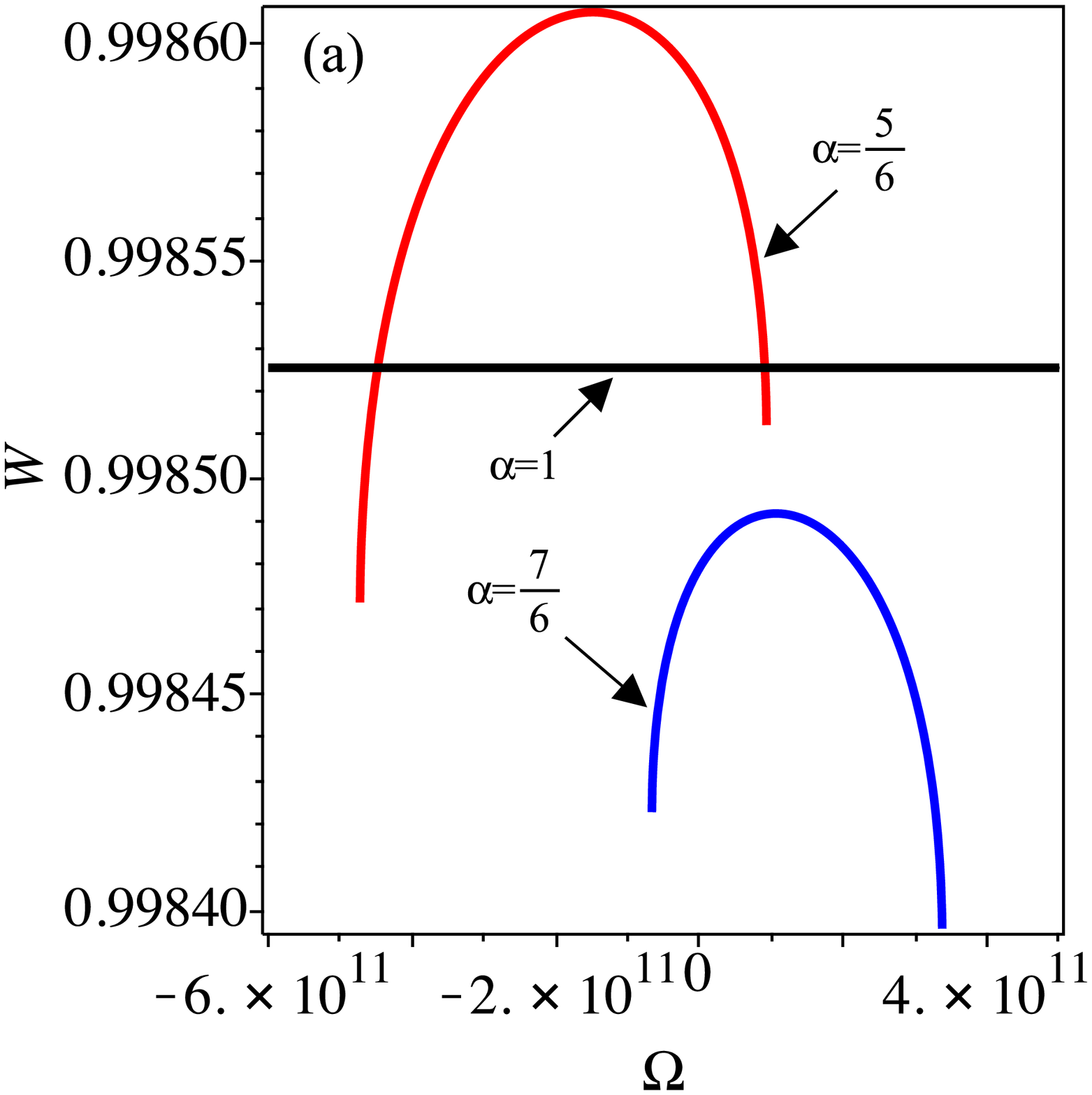}\\ \includegraphics[scale=0.35]{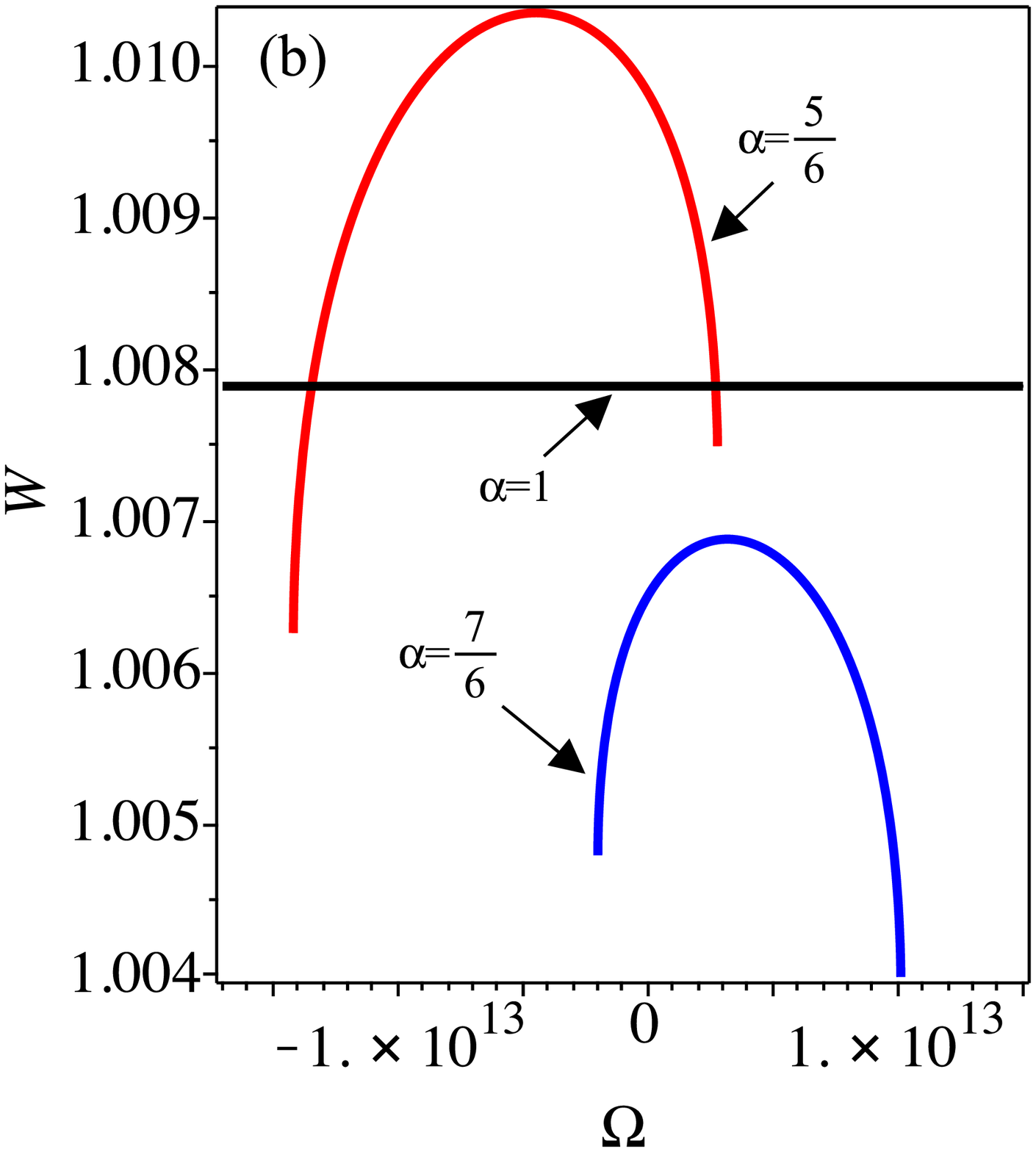}\caption{\label{fig:WvsOm}(Color online) The absorption threshold  frequency  $W$ for a quantum anti-dot as a function of angular frequency $\Omega$, for several values of the disclination parameter $\alpha$. (a) $B=0.1 T$. (b) $B=3 T$. The quantum anti-dot  size $r_0 = 895.8$ $\mathring{A}$ was assumed for all plots.}

\end{figure}

In Fig. \ref{fig:fvsr}, we plot  the absorption threshold 
frequency  $W$ as a function of the quantum anti-dot radius
$r_{0}$ for $B=1 T$.  From Fig. \ref{fig:fvsr}(a),
we see that the behavior of  $W$ is linear for $\alpha=\frac{5}{6}$
and $\alpha=1$. One interesting observation is that the presence
of the positive disclination ($\alpha=\frac{5}{6}$) shifts up the
threshold frequency. In Fig. \ref{fig:fvsr}(b) we give the plot of
$W$ as a function of radius $r_{0}$. One can be observe that when we
have negative disclination ($\alpha=\frac{7}{6}$) the behavior of
the $W$ is linear. And is always less than $\alpha=1$, when the
system has no disclination.

In Fig. \ref{fig:WvsOm}, we plot of the threshold frequency $W$
as a function of the angular frequency $\Omega$ for different values
of the magnetic fields $B$ and disclination parameter $\alpha$ for
the quantum anti-dot. As shown in Fig. \ref{fig:WvsOm}(a), the intensity
of the absorption threshold  frequency  for $\alpha=\frac{5}{6}$
is higher than in the case $\alpha=\frac{7}{6}$ in the regime of
the weak magnetic field ($B=0.1 T$). One also can see that the behavior is
different from the one shown in Fig. \ref{fig:fvsOmega}(a)  above,
since there is no crossing between the threshold frequency curves
corresponding to  the disclination parameters $\alpha=\frac{5}{6}$ and $\alpha=\frac{7}{6}$.
Finally, Fig. \ref{fig:WvsOm}(b) describes the behavior of the threshold
frequency for $B=3 T$. In this case, as  seen from the plots, the
intensity of the absorption threshold  frequency  is higher than in
the case for weak magnetic field ($B=0.1 T$).
\section{Conclusion}\label{sec3}
We have investigated the energy levels of a 2DEG in the presence of  a disclination,  under the influence of a pseudoharmonic interaction consisting of either a quantum dot or an antidot potential, in the presence of a strong uniform  magnetic field $B$, in a rotating sample. The topological defect  generates a curvature field
that acts on the particles as an external AB flux. We have  considered the effects of a covariant term, which comes from the geometric approach in the continuum limit. This is due to elastic deformations in the material with such disclination. We have found that this  changes significantly the interband light absorption in this system. Moreover, we have seen that there is a simultaneous coupling between rotation, magnetic field and the topological defect which gives rise to a range of values for the magnetic field that does not exhibit absorption of light. This is due to the fact that there is no binding of electrons in this range.  The size of this region depends on the disclination and the rotation rate. It also  depends on which material is being considered trough the effective mass of the charge carriers. The most important feature of this effect is that it only exists when the three elements (magnetic field, disclination and rotation) are present simultaneously. 

\begin{acknowledgments}
We are grateful to FAPEMIG, CNPq, CAPES and FACEPE (Brazilian agencies) for financial support. 
\end{acknowledgments}
\bibliography{2.bib}

\begin{thebibliography}{29}%
\makeatletter
\providecommand \@ifxundefined [1]{%
 \@ifx{#1\undefined}
}%
\providecommand \@ifnum [1]{%
 \ifnum #1\expandafter \@firstoftwo
 \else \expandafter \@secondoftwo
 \fi
}%
\providecommand \@ifx [1]{%
 \ifx #1\expandafter \@firstoftwo
 \else \expandafter \@secondoftwo
 \fi
}%
\providecommand \natexlab [1]{#1}%
\providecommand \enquote  [1]{``#1''}%
\providecommand \bibnamefont  [1]{#1}%
\providecommand \bibfnamefont [1]{#1}%
\providecommand \citenamefont [1]{#1}%
\providecommand \href@noop [0]{\@secondoftwo}%
\providecommand \href [0]{\begingroup \@sanitize@url \@href}%
\providecommand \@href[1]{\@@startlink{#1}\@@href}%
\providecommand \@@href[1]{\endgroup#1\@@endlink}%
\providecommand \@sanitize@url [0]{\catcode `\\12\catcode `\$12\catcode
  `\&12\catcode `\#12\catcode `\^12\catcode `\_12\catcode `\%12\relax}%
\providecommand \@@startlink[1]{}%
\providecommand \@@endlink[0]{}%
\providecommand \url  [0]{\begingroup\@sanitize@url \@url }%
\providecommand \@url [1]{\endgroup\@href {#1}{\urlprefix }}%
\providecommand \urlprefix  [0]{URL }%
\providecommand \Eprint [0]{\href }%
\providecommand \doibase [0]{http://dx.doi.org/}%
\providecommand \selectlanguage [0]{\@gobble}%
\providecommand \bibinfo  [0]{\@secondoftwo}%
\providecommand \bibfield  [0]{\@secondoftwo}%
\providecommand \translation [1]{[#1]}%
\providecommand \BibitemOpen [0]{}%
\providecommand \bibitemStop [0]{}%
\providecommand \bibitemNoStop [0]{.\EOS\space}%
\providecommand \EOS [0]{\spacefactor3000\relax}%
\providecommand \BibitemShut  [1]{\csname bibitem#1\endcsname}%
\let\auto@bib@innerbib\@empty
\bibitem [{\citenamefont {Mohammad~Bagher}(2016)}]{dotsap}%
  \BibitemOpen
  \bibfield  {author} {\bibinfo {author} {\bibfnamefont {A.}~\bibnamefont
  {Mohammad~Bagher}},\ }\href@noop {} {\bibfield  {journal} {\bibinfo
  {journal} {Sensors and Transducers}\ }\textbf {\bibinfo {volume} {198}},\
  \bibinfo {pages} {37} (\bibinfo {year} {2016})}\BibitemShut {NoStop}%
\bibitem [{\citenamefont {Nurmikko}(2015)}]{dots1}%
  \BibitemOpen
  \bibfield  {author} {\bibinfo {author} {\bibfnamefont {A.}~\bibnamefont
  {Nurmikko}},\ }\href {\doibase 10.1038/nnano.2015.288} {\bibfield  {journal}
  {\bibinfo  {journal} {Nat Nano}\ }\textbf {\bibinfo {volume} {10}},\ \bibinfo
  {pages} {1001} (\bibinfo {year} {2015})}\BibitemShut {NoStop}%
\bibitem [{\citenamefont {Tan}\ and\ \citenamefont
  {Inkson}(1996{\natexlab{a}})}]{PRB.1996.53.6947}%
  \BibitemOpen
  \bibfield  {author} {\bibinfo {author} {\bibfnamefont {W.-C.}\ \bibnamefont
  {Tan}}\ and\ \bibinfo {author} {\bibfnamefont {J.~C.}\ \bibnamefont
  {Inkson}},\ }\href {\doibase 10.1103/PhysRevB.53.6947} {\bibfield  {journal}
  {\bibinfo  {journal} {Phys. Rev. B}\ }\textbf {\bibinfo {volume} {53}},\
  \bibinfo {pages} {6947} (\bibinfo {year} {1996}{\natexlab{a}})}\BibitemShut
  {NoStop}%
\bibitem [{\citenamefont {Katanaev}(2005)}]{1063-7869-48-7-R02}%
  \BibitemOpen
  \bibfield  {author} {\bibinfo {author} {\bibfnamefont {M.~O.}\ \bibnamefont
  {Katanaev}},\ }\href@noop {} {\bibfield  {journal} {\bibinfo  {journal}
  {Physics-Uspekhi}\ }\textbf {\bibinfo {volume} {48}},\ \bibinfo {pages} {675}
  (\bibinfo {year} {2005})}\BibitemShut {NoStop}%
\bibitem [{\citenamefont {de~Lima}\ and\ \citenamefont
  {Filgueiras}(2012{\natexlab{a}})}]{de2012integer}%
  \BibitemOpen
  \bibfield  {author} {\bibinfo {author} {\bibfnamefont {A.}~\bibnamefont
  {de~Lima}}\ and\ \bibinfo {author} {\bibfnamefont {C.}~\bibnamefont
  {Filgueiras}},\ }\href@noop {} {\bibfield  {journal} {\bibinfo  {journal}
  {The European Physical Journal B}\ }\textbf {\bibinfo {volume} {85}},\
  \bibinfo {pages} {401} (\bibinfo {year} {2012}{\natexlab{a}})}\BibitemShut
  {NoStop}%
\bibitem [{\citenamefont {de~Lima}\ \emph {et~al.}(2013)\citenamefont
  {de~Lima}, \citenamefont {Poux}, \citenamefont {Assafr{\~a}o},\ and\
  \citenamefont {Filgueiras}}]{de2013screw}%
  \BibitemOpen
  \bibfield  {author} {\bibinfo {author} {\bibfnamefont {A.~G.}\ \bibnamefont
  {de~Lima}}, \bibinfo {author} {\bibfnamefont {A.}~\bibnamefont {Poux}},
  \bibinfo {author} {\bibfnamefont {D.}~\bibnamefont {Assafr{\~a}o}}, \ and\
  \bibinfo {author} {\bibfnamefont {C.}~\bibnamefont {Filgueiras}},\
  }\href@noop {} {\bibfield  {journal} {\bibinfo  {journal} {The European
  Physical Journal B}\ }\textbf {\bibinfo {volume} {86}},\ \bibinfo {pages} {1}
  (\bibinfo {year} {2013})}\BibitemShut {NoStop}%
\bibitem [{\citenamefont {Bakke}\ and\ \citenamefont
  {Moraes}(2012)}]{PLA.2012.376.2838}%
  \BibitemOpen
  \bibfield  {author} {\bibinfo {author} {\bibfnamefont {K.}~\bibnamefont
  {Bakke}}\ and\ \bibinfo {author} {\bibfnamefont {F.}~\bibnamefont {Moraes}},\
  }\href {\doibase http://dx.doi.org/10.1016/j.physleta.2012.09.006} {\bibfield
   {journal} {\bibinfo  {journal} {Phys. Lett. A}\ }\textbf {\bibinfo {volume}
  {376}},\ \bibinfo {pages} {2838 } (\bibinfo {year} {2012})}\BibitemShut
  {NoStop}%
\bibitem [{\citenamefont {Fumeron}\ \emph {et~al.}(2017)\citenamefont
  {Fumeron}, \citenamefont {Berche}, \citenamefont {Medina}, \citenamefont
  {Santos},\ and\ \citenamefont {Moraes}}]{fumeron2017using}%
  \BibitemOpen
  \bibfield  {author} {\bibinfo {author} {\bibfnamefont {S.}~\bibnamefont
  {Fumeron}}, \bibinfo {author} {\bibfnamefont {B.}~\bibnamefont {Berche}},
  \bibinfo {author} {\bibfnamefont {E.}~\bibnamefont {Medina}}, \bibinfo
  {author} {\bibfnamefont {F.~A.}\ \bibnamefont {Santos}}, \ and\ \bibinfo
  {author} {\bibfnamefont {F.}~\bibnamefont {Moraes}},\ }\href@noop {}
  {\bibfield  {journal} {\bibinfo  {journal} {EPL (Europhysics Letters)}\
  }\textbf {\bibinfo {volume} {117}},\ \bibinfo {pages} {47007} (\bibinfo
  {year} {2017})}\BibitemShut {NoStop}%
\bibitem [{\citenamefont {Filgueiras}\ \emph {et~al.}(2016)\citenamefont
  {Filgueiras}, \citenamefont {Rojas}, \citenamefont {Aciole},\ and\
  \citenamefont {Silva}}]{filgueiras2016landau}%
  \BibitemOpen
  \bibfield  {author} {\bibinfo {author} {\bibfnamefont {C.}~\bibnamefont
  {Filgueiras}}, \bibinfo {author} {\bibfnamefont {M.}~\bibnamefont {Rojas}},
  \bibinfo {author} {\bibfnamefont {G.}~\bibnamefont {Aciole}}, \ and\ \bibinfo
  {author} {\bibfnamefont {E.~O.}\ \bibnamefont {Silva}},\ }\href@noop {}
  {\bibfield  {journal} {\bibinfo  {journal} {Physics Letters A}\ }\textbf
  {\bibinfo {volume} {380}},\ \bibinfo {pages} {3847} (\bibinfo {year}
  {2016})}\BibitemShut {NoStop}%
\bibitem [{\citenamefont {Barnett}(1915)}]{barnett1915magnetization}%
  \BibitemOpen
  \bibfield  {author} {\bibinfo {author} {\bibfnamefont {S.~J.}\ \bibnamefont
  {Barnett}},\ }\href@noop {} {\bibfield  {journal} {\bibinfo  {journal}
  {Physical Review}\ }\textbf {\bibinfo {volume} {6}},\ \bibinfo {pages} {239}
  (\bibinfo {year} {1915})}\BibitemShut {NoStop}%
\bibitem [{\citenamefont {Lima}\ \emph {et~al.}(2014)\citenamefont {Lima},
  \citenamefont {Brand{\~a}o}, \citenamefont {Cunha},\ and\ \citenamefont
  {Moraes}}]{Lima2014}%
  \BibitemOpen
  \bibfield  {author} {\bibinfo {author} {\bibfnamefont {J.~R.~F.}\
  \bibnamefont {Lima}}, \bibinfo {author} {\bibfnamefont {J.}~\bibnamefont
  {Brand{\~a}o}}, \bibinfo {author} {\bibfnamefont {M.~M.}\ \bibnamefont
  {Cunha}}, \ and\ \bibinfo {author} {\bibfnamefont {F.}~\bibnamefont
  {Moraes}},\ }\href {\doibase 10.1140/epjd/e2014-40570-4} {\bibfield
  {journal} {\bibinfo  {journal} {The European Physical Journal D}\ }\textbf
  {\bibinfo {volume} {68}},\ \bibinfo {pages} {94} (\bibinfo {year}
  {2014})}\BibitemShut {NoStop}%
\bibitem [{\citenamefont {Lima}\ and\ \citenamefont {Moraes}(2015)}]{Lima2015}%
  \BibitemOpen
  \bibfield  {author} {\bibinfo {author} {\bibfnamefont {J.~R.~F.}\
  \bibnamefont {Lima}}\ and\ \bibinfo {author} {\bibfnamefont {F.}~\bibnamefont
  {Moraes}},\ }\href {\doibase 10.1140/epjb/e2015-60022-9} {\bibfield
  {journal} {\bibinfo  {journal} {The European Physical Journal B}\ }\textbf
  {\bibinfo {volume} {88}},\ \bibinfo {pages} {63} (\bibinfo {year}
  {2015})}\BibitemShut {NoStop}%
\bibitem [{\citenamefont {Fonseca}\ and\ \citenamefont
  {Bakke}(2016)}]{knutrotation}%
  \BibitemOpen
  \bibfield  {author} {\bibinfo {author} {\bibfnamefont {I.~C.}\ \bibnamefont
  {Fonseca}}\ and\ \bibinfo {author} {\bibfnamefont {K.}~\bibnamefont
  {Bakke}},\ }\href {\doibase 10.1063/1.4939525} {\bibfield  {journal}
  {\bibinfo  {journal} {The Journal of Chemical Physics}\ }\textbf {\bibinfo
  {volume} {144}},\ \bibinfo {pages} {014308} (\bibinfo {year}
  {2016})}\BibitemShut {NoStop}%
\bibitem [{\citenamefont {Dayi}\ \emph {et~al.}(2017)\citenamefont {Dayi},
  \citenamefont {Kilin\ifmmode~\mbox{\c{c}}\else \c{c}\fi{}arslan},\ and\
  \citenamefont {Yunt}}]{PhysRevD.95.085005}%
  \BibitemOpen
  \bibfield  {author} {\bibinfo {author} {\bibfnamefont {O.~F.}\ \bibnamefont
  {Dayi}}, \bibinfo {author} {\bibfnamefont {E.}~\bibnamefont
  {Kilin\ifmmode~\mbox{\c{c}}\else \c{c}\fi{}arslan}}, \ and\ \bibinfo {author}
  {\bibfnamefont {E.}~\bibnamefont {Yunt}},\ }\href {\doibase
  10.1103/PhysRevD.95.085005} {\bibfield  {journal} {\bibinfo  {journal} {Phys.
  Rev. D}\ }\textbf {\bibinfo {volume} {95}},\ \bibinfo {pages} {085005}
  (\bibinfo {year} {2017})}\BibitemShut {NoStop}%
\bibitem [{\citenamefont {Filgueiras}\ \emph {et~al.}(2015)\citenamefont
  {Filgueiras}, \citenamefont {Brandao},\ and\ \citenamefont
  {Moraes}}]{EPLrotation}%
  \BibitemOpen
  \bibfield  {author} {\bibinfo {author} {\bibfnamefont {C.}~\bibnamefont
  {Filgueiras}}, \bibinfo {author} {\bibfnamefont {J.}~\bibnamefont {Brandao}},
  \ and\ \bibinfo {author} {\bibfnamefont {F.}~\bibnamefont {Moraes}},\ }\href
  {http://stacks.iop.org/0295-5075/110/i=2/a=27003} {\bibfield  {journal}
  {\bibinfo  {journal} {EPL (Europhysics Letters)}\ }\textbf {\bibinfo {volume}
  {110}},\ \bibinfo {pages} {27003} (\bibinfo {year} {2015})}\BibitemShut
  {NoStop}%
\bibitem [{\citenamefont {Brandao}\ \emph {et~al.}(2015)\citenamefont
  {Brandao}, \citenamefont {Moraes}, \citenamefont {Cunha}, \citenamefont
  {Lima},\ and\ \citenamefont {Filgueiras}}]{Brandao201555}%
  \BibitemOpen
  \bibfield  {author} {\bibinfo {author} {\bibfnamefont {J.~E.}\ \bibnamefont
  {Brandao}}, \bibinfo {author} {\bibfnamefont {F.}~\bibnamefont {Moraes}},
  \bibinfo {author} {\bibfnamefont {M.}~\bibnamefont {Cunha}}, \bibinfo
  {author} {\bibfnamefont {J.~R.}\ \bibnamefont {Lima}}, \ and\ \bibinfo
  {author} {\bibfnamefont {C.}~\bibnamefont {Filgueiras}},\ }\href {\doibase
  https://doi.org/10.1016/j.rinp.2015.02.003} {\bibfield  {journal} {\bibinfo
  {journal} {Results in Physics}\ }\textbf {\bibinfo {volume} {5}},\ \bibinfo
  {pages} {55 } (\bibinfo {year} {2015})}\BibitemShut {NoStop}%
\bibitem [{\citenamefont {S\'atiro}\ \emph {et~al.}(2009)\citenamefont
  {S\'atiro}, \citenamefont {de~M.~Carvalho},\ and\ \citenamefont
  {Moraes}}]{doi:10.1142/S0217732309029995}%
  \BibitemOpen
  \bibfield  {author} {\bibinfo {author} {\bibfnamefont {C.}~\bibnamefont
  {S\'atiro}}, \bibinfo {author} {\bibfnamefont {A.~M.}\ \bibnamefont
  {de~M.~Carvalho}}, \ and\ \bibinfo {author} {\bibfnamefont {F.}~\bibnamefont
  {Moraes}},\ }\href {\doibase 10.1142/S0217732309029995} {\bibfield  {journal}
  {\bibinfo  {journal} {Modern Physics Letters A}\ }\textbf {\bibinfo {volume}
  {24}},\ \bibinfo {pages} {1437} (\bibinfo {year} {2009})}\BibitemShut
  {NoStop}%
\bibitem [{\citenamefont {Ikhdair}\ and\ \citenamefont
  {Hamzavi}(2012)}]{PB.2012.407.4198}%
  \BibitemOpen
  \bibfield  {author} {\bibinfo {author} {\bibfnamefont {S.~M.}\ \bibnamefont
  {Ikhdair}}\ and\ \bibinfo {author} {\bibfnamefont {M.}~\bibnamefont
  {Hamzavi}},\ }\href {\doibase http://dx.doi.org/10.1016/j.physb.2012.07.004}
  {\bibfield  {journal} {\bibinfo  {journal} {Physica B: Condensed Matter}\
  }\textbf {\bibinfo {volume} {407}},\ \bibinfo {pages} {4198 } (\bibinfo
  {year} {2012})}\BibitemShut {NoStop}%
\bibitem [{\citenamefont {Tan}\ and\ \citenamefont
  {Inkson}(1996{\natexlab{b}})}]{SST.1996.11.1635}%
  \BibitemOpen
  \bibfield  {author} {\bibinfo {author} {\bibfnamefont {W.-C.}\ \bibnamefont
  {Tan}}\ and\ \bibinfo {author} {\bibfnamefont {J.~C.}\ \bibnamefont
  {Inkson}},\ }\href {\doibase 10.1088/0268-1242/11/11/001} {\bibfield
  {journal} {\bibinfo  {journal} {Semicond. Sci. Technol.}\ }\textbf {\bibinfo
  {volume} {11}},\ \bibinfo {pages} {1635} (\bibinfo {year}
  {1996}{\natexlab{b}})}\BibitemShut {NoStop}%
\bibitem [{\citenamefont {Abramowitz}\ and\ \citenamefont
  {Stegun}(1972)}]{Book.1972.Abramowitz}%
  \BibitemOpen
  \bibinfo {editor} {\bibfnamefont {M.}~\bibnamefont {Abramowitz}}\ and\
  \bibinfo {editor} {\bibfnamefont {I.~A.}\ \bibnamefont {Stegun}},\ eds.,\
  \href@noop {} {\emph {\bibinfo {title} {Handbook of Mathematical
  Functions}}}\ (\bibinfo  {publisher} {New York: Dover Publications},\
  \bibinfo {year} {1972})\BibitemShut {NoStop}%
\bibitem [{\citenamefont {Jensen}\ and\ \citenamefont
  {Dandoloff}(2011)}]{Jensen2011448}%
  \BibitemOpen
  \bibfield  {author} {\bibinfo {author} {\bibfnamefont {B.}~\bibnamefont
  {Jensen}}\ and\ \bibinfo {author} {\bibfnamefont {R.}~\bibnamefont
  {Dandoloff}},\ }\href {\doibase
  http://dx.doi.org/10.1016/j.physleta.2010.12.018} {\bibfield  {journal}
  {\bibinfo  {journal} {Physics Letters A}\ }\textbf {\bibinfo {volume}
  {375}},\ \bibinfo {pages} {448 } (\bibinfo {year} {2011})}\BibitemShut
  {NoStop}%
\bibitem [{\citenamefont {Andrade}\ \emph {et~al.}(2013)\citenamefont
  {Andrade}, \citenamefont {Silva},\ and\ \citenamefont
  {Pereira}}]{AoP.2013.339.510}%
  \BibitemOpen
  \bibfield  {author} {\bibinfo {author} {\bibfnamefont {F.~M.}\ \bibnamefont
  {Andrade}}, \bibinfo {author} {\bibfnamefont {E.~O.}\ \bibnamefont {Silva}},
  \ and\ \bibinfo {author} {\bibfnamefont {M.}~\bibnamefont {Pereira}},\ }\href
  {\doibase 10.1016/j.aop.2013.10.001} {\bibfield  {journal} {\bibinfo
  {journal} {Ann. Phys. (NY)}\ }\textbf {\bibinfo {volume} {339}},\ \bibinfo
  {pages} {510} (\bibinfo {year} {2013})}\BibitemShut {NoStop}%
\bibitem [{\citenamefont {Khalilov}\ and\ \citenamefont
  {Mamsurov}(2009)}]{TMP.2009.161.1503}%
  \BibitemOpen
  \bibfield  {author} {\bibinfo {author} {\bibfnamefont {V.}~\bibnamefont
  {Khalilov}}\ and\ \bibinfo {author} {\bibfnamefont {I.}~\bibnamefont
  {Mamsurov}},\ }\href {\doibase 10.1007/s11232-009-0137-9} {\bibfield
  {journal} {\bibinfo  {journal} {Theor. Math. Phys.}\ }\textbf {\bibinfo
  {volume} {161}},\ \bibinfo {pages} {1503} (\bibinfo {year}
  {2009})}\BibitemShut {NoStop}%
\bibitem [{\citenamefont {Khalilov}(2014)}]{EPJC.2014.74.2708}%
  \BibitemOpen
  \bibfield  {author} {\bibinfo {author} {\bibfnamefont {V.}~\bibnamefont
  {Khalilov}},\ }\href {\doibase 10.1140/epjc/s10052-013-2708-z} {\bibfield
  {journal} {\bibinfo  {journal} {European Physical Journal C}\ }\textbf
  {\bibinfo {volume} {74}},\ \bibinfo {pages} {1} (\bibinfo {year}
  {2014})}\BibitemShut {NoStop}%
\bibitem [{\citenamefont {de~Lima}\ and\ \citenamefont
  {Filgueiras}(2012{\natexlab{b}})}]{deLima2012}%
  \BibitemOpen
  \bibfield  {author} {\bibinfo {author} {\bibfnamefont {A.}~\bibnamefont
  {de~Lima}}\ and\ \bibinfo {author} {\bibfnamefont {C.}~\bibnamefont
  {Filgueiras}},\ }\href {\doibase 10.1140/epjb/e2012-30766-9} {\bibfield
  {journal} {\bibinfo  {journal} {The European Physical Journal B}\ }\textbf
  {\bibinfo {volume} {85}},\ \bibinfo {pages} {401} (\bibinfo {year}
  {2012}{\natexlab{b}})}\BibitemShut {NoStop}%
\bibitem [{\citenamefont {Atoyan}\ \emph {et~al.}(2004)\citenamefont {Atoyan},
  \citenamefont {Kazaryan},\ and\ \citenamefont {Sarkisyan}}]{PE.2004.22.860}%
  \BibitemOpen
  \bibfield  {author} {\bibinfo {author} {\bibfnamefont {M.}~\bibnamefont
  {Atoyan}}, \bibinfo {author} {\bibfnamefont {E.}~\bibnamefont {Kazaryan}}, \
  and\ \bibinfo {author} {\bibfnamefont {H.}~\bibnamefont {Sarkisyan}},\ }\href
  {\doibase http://dx.doi.org/10.1016/j.physe.2003.09.042} {\bibfield
  {journal} {\bibinfo  {journal} {Physica E: Low-dimensional Systems and
  Nanostructures}\ }\textbf {\bibinfo {volume} {22}},\ \bibinfo {pages} {860 }
  (\bibinfo {year} {2004})}\BibitemShut {NoStop}%
\bibitem [{\citenamefont {Atoyan}\ \emph {et~al.}(2006)\citenamefont {Atoyan},
  \citenamefont {Kazaryan},\ and\ \citenamefont {Sarkisyan}}]{PE.2006.31.83}%
  \BibitemOpen
  \bibfield  {author} {\bibinfo {author} {\bibfnamefont {M.}~\bibnamefont
  {Atoyan}}, \bibinfo {author} {\bibfnamefont {E.}~\bibnamefont {Kazaryan}}, \
  and\ \bibinfo {author} {\bibfnamefont {H.}~\bibnamefont {Sarkisyan}},\ }\href
  {\doibase http://dx.doi.org/10.1016/j.physe.2005.10.008} {\bibfield
  {journal} {\bibinfo  {journal} {Physica E: Low-dimensional Systems and
  Nanostructures}\ }\textbf {\bibinfo {volume} {31}},\ \bibinfo {pages} {83 }
  (\bibinfo {year} {2006})}\BibitemShut {NoStop}%
\bibitem [{\citenamefont {Raigoza}\ \emph {et~al.}(2005)\citenamefont
  {Raigoza}, \citenamefont {Morales},\ and\ \citenamefont
  {Duque}}]{PB.2005.363.262}%
  \BibitemOpen
  \bibfield  {author} {\bibinfo {author} {\bibfnamefont {N.}~\bibnamefont
  {Raigoza}}, \bibinfo {author} {\bibfnamefont {A.}~\bibnamefont {Morales}}, \
  and\ \bibinfo {author} {\bibfnamefont {C.}~\bibnamefont {Duque}},\ }\href
  {\doibase http://dx.doi.org/10.1016/j.physb.2005.03.031} {\bibfield
  {journal} {\bibinfo  {journal} {Physica B: Condensed Matter}\ }\textbf
  {\bibinfo {volume} {363}},\ \bibinfo {pages} {262 } (\bibinfo {year}
  {2005})}\BibitemShut {NoStop}%
\bibitem [{\citenamefont {Khordad}(2010)}]{SSS.2010.12.1253}%
  \BibitemOpen
  \bibfield  {author} {\bibinfo {author} {\bibfnamefont {R.}~\bibnamefont
  {Khordad}},\ }\href {\doibase
  http://dx.doi.org/10.1016/j.solidstatesciences.2010.03.001} {\bibfield
  {journal} {\bibinfo  {journal} {Solid State Sciences}\ }\textbf {\bibinfo
  {volume} {12}},\ \bibinfo {pages} {1253 } (\bibinfo {year}
  {2010})}\BibitemShut {NoStop}%
\end{thebibliography}%
\end{document}